\documentclass[useAMS,usenatbib]{mn2e}

\usepackage{amsmath, amssymb}
\usepackage{eucal}          
\usepackage{epsfig}         
\usepackage{graphicx}
\usepackage{float, stfloats}
\usepackage{multirow}
\usepackage{tabularx}

\usepackage{natbib}
\usepackage{aas_macros}

\usepackage{url}
\usepackage[draft]{hyperref}

\newcolumntype{L}[1]{>{\raggedright\arraybackslash}p{#1}}
\newcolumntype{C}[1]{>{\centering\arraybackslash}p{#1}}
\newcolumntype{R}[1]{>{\raggedleft\arraybackslash}p{#1}}
\usepackage{xcolor,colortbl}
\definecolor{Gray}{gray}{0.85}
\newcolumntype{G}{>{\columncolor{Gray}}r}

\def\Mpch{~h^{-1} {\rm Mpc}}
\def\MpchVolume{~(h^{-1} {\rm Mpc})^3}
\def\kpc{~\rm kpc}
\def\kpch{~h^{-1} {\rm kpc}}
\def\sun{\odot}
\def\Msun{\rm{M}_{\odot}}
\def\Msolar{~h^{-1} \rm{M}_{\odot}}

\def\kms{~\rm{km/s}}

\def\marcsyear{~\rm{mas\;yr^{-1}}}

\newcommand{\MII}{\textsc{MS-II}}
\newcommand{\COCO}{\textsc{COCO}}

\newcommand {\lcdm}{$\Lambda$CDM}

\newcommand {\pOne}  {\emph{spatial}}
\newcommand {\pTwo}  {\emph{spatial + 2D-kinematic}}
\newcommand {\pThree}{\emph{spatial + 3D-kinematic}}
\newcommand {\pFour} {\emph{2D-kinematic}}
\newcommand {\pFive} {\emph{3D-kinematic}}
\newcommand {\plOne}  {spatial}
\newcommand {\plTwo}  {spatial + 2D-kinematic}
\newcommand {\plThree}{spatial + 3D-kinematic}
\newcommand {\planeOne}{\emph{spatial}}
\newcommand {\planeTwo}{\emph{spatial + 2D-kinematic}}
\newcommand {\planeThree}{\emph{spatial + 3D-kinematic}}

\newcommand{\Prominence}{\mathcal{P}}
\newcommand{\Prarest}{\Prominence^{\,\rm rarest}}
\newcommand{\Pspatial}   {\Prarest_\textrm{spatial}}

\newcommand{\Pcombined}  {\Prarest_\textrm{spatial\,+\,2D-kin}}
\newcommand{\PcombinedMW}{\Prarest_\textrm{spatial\,+\,3D-kin}}

\newcommand{\Nmax}{N_{\rm max}}
\newcommand{\Nsat}{N_{\rm sat}}
\newcommand{\senseRot}{s.s.r.}
\newcommand{\Ncor}{N_{\rm \senseRot{}}}
\newcommand{\rPerp}{r_{\perp}}
\newcommand{\rAlong}{r_{\parallel}}
\newcommand{\Dorbit}{\Delta_{\rm std}}

\newcommand{\fLCDM}{f_{\rm \Lambda CDM}}
\newcommand{\fractLCDM}{f_{\rm \Lambda CDM}(\Nsat, \allowbreak {\le}\rPerp ,\allowbreak {\ge}\rAlong, \allowbreak {\ge}\Ncor)}
\newcommand{\fractLCDMorbit}{f_{\rm \Lambda CDM}(\Nsat, \allowbreak {\le}\rPerp, \allowbreak {\ge}\rAlong, \allowbreak {\le}\Dorbit)}

\newcommand{\eq}[1]{Eq. \eqref{#1}}

\newcommand{\refsec}[1]{Sec.~\ref{#1}}
\newcommand{\refsecs}[2]{Secs.~\ref{#1} and \ref{#2}}
\newcommand{\refappendix}[1]{Appendix~\ref{#1}}
\newcommand{\reftab}[1]{Table~\ref{#1}}
\newcommand{\reffig}[1]{Fig.~\ref{#1}}
\newcommand{\reffigs}[2]{Figs.~\ref{#1}-\ref{#2}}
\newcommand{\reffigS}[2]{Figs.~\ref{#1} and \ref{#2}}

\newcommand{\figDir}{fig_pdf/}

\newcommand{\pandas}{PAndAS}

\newcommand{\ibata}{\hyperlink{labelHypertarget}{Ibata13}}

\usepackage[normalem]{ulem}
\usepackage{color}
\definecolor{colorChanges}{rgb}{.0,.3,1.}

\newcommand{\MCn}[1]{#1} 
\newcommand{\MCd}[1]{} 
\newcommand{\MCc}[1]{} 
\newcommand{\MCq}[1]{} 

\title{Planes of satellite galaxies: when exceptions are the rule}

\author[Cautun et~al.]
{\parbox{\textwidth}{
        Marius Cautun$^{1}$\thanks{E-mail : m.c.cautun@durham.ac.uk},
        Sownak Bose$^{1}$,
        Carlos~S.~Frenk$^{1}$,
        Qi Guo$^{2}$,
        Jiaxin Han$^{1}$,  \\
        Wojciech A. Hellwing$^{1,3}$,
        Till Sawala$^{1}$
        and Wenting Wang$^{1}$\vspace{.3cm}
        } \\
$^{1}$Institute for Computational Cosmology, Department of Physics, Durham University, South Road, Durham DH1 3LE, UK\\
$^{2}$Key Laboratory for Computational Astrophysics, The Partner Group of Max Planck Institute for Astrophysics, National Astronomical\\ Observatories, Chinese Academy of Sciences, Beijing, 100012, China \\
$^{3}$Interdisciplinary Centre for Mathematical and Computational Modelling, University of Warsaw, ul. Pawi\'nskiego 5a, Warsaw, Poland\\
}

\begin{document}


\maketitle

\begin{abstract}
    The detection of planar structures within the satellite systems of both the Milky Way (MW) and Andromeda (M31) has been reported as being in stark contradiction to the predictions of the standard cosmological model (\lcdm{}). Given the ambiguity in defining a planar configuration, it is unclear how to interpret the low incidence of the MW and M31 planes in \lcdm{}. We investigate the prevalence of satellite planes around galactic mass haloes identified in high resolution cosmological simulations. We find that planar structures are very common, and that ${\sim}10\%$ of \lcdm{} haloes have even more prominent planes than those present in the Local Group. While ubiquitous, the planes of satellite galaxies show a large diversity in their properties. This precludes using one or two systems as small scale probes of cosmology, since a large sample of satellite systems is needed to obtain a good measure of the object-to-object variation. This very diversity has been misinterpreted as a discrepancy between the satellite planes observed in the Local Group and \lcdm{} predictions. In fact, ${\sim}10\%$ of \lcdm{} galactic haloes have planes of satellites that are as infrequent as the MW and M31 planes. The \emph{look-elsewhere effect} plays an important role in assessing the detection significance of satellite planes and accounting for it leads to overestimating the significance level by a factor of 30 and 100 for the MW and M31 systems, respectively. 
\end{abstract} 



\section{Introduction}
\label{sec:introduction}
While the Universe at large may be homogeneous and isotropic, on galactic scale the distribution of galaxies is highly anisotropic. This is most readily seen in the spatial and kinematical distribution of the Local Group (LG) satellites. In the MW, the 11 ``classical'' satellites define a thin plane \citep{Lynden-Bell1976} and some of the fainter satellites, tidal streams and young globular clusters have an anisotropic distribution reminiscent of this plane \citep{Metz2009,Pawlowski2012}. 
Many members of this ``disk of
satellites'' have a common rotation direction and it has been claimed
that the plane is a rotationally stabilized structure
\citep{Metz2008c,Pawlowski2013b}. Similarly, the spatial distribution
of satellites around M31 is anisotropic
\citep{Koch2006,McConnachie2006}, with 15 out of 27 satellites
observed by the Pan-Andromeda Archaeological
Survey\citep[\pandas{};][]{McConnachie2009} located in a thin plane.
Out of the 15 members of the plane, 13 of them share the same sense of rotation
\citep[][hereafter \ibata{}]{Ibata2013}.

Anisotropies in the distribution of satellites are a clear prediction
of the $\Lambda$ cold dark matter (\lcdm) paradigm
\citep{Libeskind2005,Zentner2005a,Libeskind2009,Libeskind2011,Deason2011,Wang2013,Sawala2014c}.
Such flattened satellite distributions, dubbed ``great pancakes'', can
arise from the infall of satellites along the spine of filaments
\citep{Libeskind2005,Buck2015}, which in turn determine the preferential points
at which satellites enter the virial radius of the host halo
\citep{Libeskind2011,Libeskind2014}. The imprint of anisotropic accretion is retained in
the dynamics of satellites, with a significant population
co-rotating with the spin of the host halo \citep{Libeskind2009,Lovell2011,Cautun2015a}.

Although flattened satellite distributions are common in \lcdm{}, configurations
similar to those of the MW and M31 are infrequent. 
\citet{Wang2013} found that $5-10\%$ of satellite systems
are as flat as the MW's 11 classical satellites. When it is required
that the velocities of at least 8 of the 11 satellites
should point within the narrow angle claimed by \citet{Pawlowski2013b}
for the MW satellites, this fraction decreases to ${\sim}0.1\%$ \citep{Pawlowski2014c}. 
In the case of the M31 thin satellite plane, \citet{Bahl2014} found that, while similar spatial distributions of
satellites are quite common in \lcdm{}, there is only a $2\%$ chance
that 13 out of the 15 members in the plane would share the same sense of
rotation. In similar studies, \citet{Ibata2014b} and \citet{Pawlowski2014c} found an even lower occurrence for 
the M31 plane, with only ${\sim}0.1\%$ of \lcdm{} systems having a similar configuration.

Extending the above analysis to galaxies outside the LG is constrained by observational limitations, but some additional tests can be performed. \citet{Cautun2015a} studied the flattening of the satellite distribution around isolated central galaxies in the SDSS, as viewed on the plane of the sky, finding good agreement between data and cosmological simulations. Using a similar approach, \citet{Ibata2014d} claimed a higher degree of flattening in their data, but their conclusions may be affected by systematics \citep[see][for a discussion of this study]{Cautun2015a}. The expected signature of planar rotation has been investigated by considering the velocity correlation of satellite pairs observed on opposite sides of the host galaxy. Initially, using a sample of 23 systems, \citet{Ibata2014} reported a significant excess, when compared to \lcdm{} predictions. \citet[][see also \citealt{Phillips2015}]{Cautun2015a} found that this excess decreases rapidly as the sample size is increased and also that the expected mirror image signal is absent for satellite pairs on the same side of the host galaxy, suggesting that the claimed excess is not robust. 

In this study, we examine planar configurations of satellites identified in MW- and M31-like mock \lcdm{} catalogues and compare them to the planar structures observed in the LG. Among others, we revisit the claims by \citet{Ibata2014b} and \citet{Pawlowski2014c} that the two planes of satellites found in the LG are inconsistent with \lcdm{} predictions. Along with similar studies,
those works are based on two important axioms: that the majority of planar configurations are the same, and that the planes detected around the MW and M31 are representative of planar structures in general. We will examine these two conjectures within the context of \lcdm{} predictions. Starting from high resolution cosmological simulations, we create mock catalogues that are used to identify planar satellite configurations similar to those found in the LG. We proceed to study the properties of the most prominent planes of satellites and to compare those \lcdm{} predictions with the two planes of satellites observed in MW and M31.

This paper is organized as follows. In \refsec{sec:data} we introduce the cosmological simulations as well as the selection criteria used to identify MW and M31 analogue systems; in \refsec{sec:defining_planes} we present an objective method to identify spatially and kinematically rare planes; in \refsec{sec:M31_planes} we identify planes of satellites for \pandas{}-like mocks; in \refsec{sec:MW_planes} we analyse MW-like systems; we conclude with a short discussion and summary in \refsec{sec:conclusion}.


\section{Data and sample selection}
\label{sec:data}
This work makes use of two high-resolution \lcdm{} DM-only cosmological simulations: the Millennium-II \citep[\MII{};][]{Boylan-Kolchin2009} and the {\it Copernicus Complexio} \citep[\COCO{};][]{Hellwing2015} simulations. Instead of using the original \MII{}, which was run assuming a \textit{Wilkinson Microwave Anisotropy Probe} (WMAP)-1 cosmogony, we employ a modified version of the simulation that has been rescaled to the WMAP-7 cosmology ($\Omega_m=0.272$, $\Omega_{\Lambda}=0.728$, $\sigma_8=0.81$ and $n_s=0.968$) using the scaling algorithm of \citet[][see also \citealt{Guo2013}]{Angulo2010}. The rescaled \MII{} corresponds to a simulation in a periodic box of side-length $104.3\Mpch$ containing $2160^3$ DM particles, each particle having a mass, $m_p=8.50\times10^{6}\Msolar$. 

\COCO{} simulates a smaller, \MCn{roughly spherical}, cosmological volume, $V = 2.25\times 10^4 \MpchVolume$, \MCn{equivalent to a cubic box of side-length $28.2\Mpch$}, but at much higher resolution than \MII{}, having $2374^3$ particles each with mass, $m_p=1.14\times10^5\Msolar$, and a gravitational softening length, $\epsilon=0.23\kpch$. This volume is embedded within a larger periodic box, of side-length $70.4\Mpch$, that is resolved at a significantly lower resolution \citep[see ][for more details]{Hellwing2015}. \COCO{} uses the same WMAP-7 cosmological parameters as the rescaled \MII{}.

To construct mock catalogues, we use the semi-analytic galaxy formation model of \citet{Guo2011_SAM} that has been implemented for the rescaled version of \MII{} in \citet{Guo2013}  and for \COCO{} in \citet{Guo2015}. The semi-analytic model has been calibrated to reproduce the stellar mass, luminosity and autocorrelation functions of low redshift galaxies as inferred from SDSS. The abundance and radial distribution of satellite galaxies predicted by the model are in good agreement with SDSS data \citep{Wang2012,WangW2014}, which makes the \citeauthor{Guo2011_SAM} model a good test bed for studying planar structures of satellites.

Due to the relatively low resolution of \MII{}, many of the satellite galaxies of interest for this work are found in haloes close to the resolution limit of ${\sim}10^{9}\Msolar$, which raises questions about the accuracy of the properties and orbits of these objects, especially after infall into the main halo. 
To test for any potential systematic effects arising from the limited resolution of \MII{}, we compare with the results of \COCO{}. Any such effects are significantly reduced or even absent in \COCO{}, since it has $75$ times higher mass resolution and four times better spatial resolution.

\begin{figure*}
     \centering
     \includegraphics[width=0.3\linewidth,angle=0]{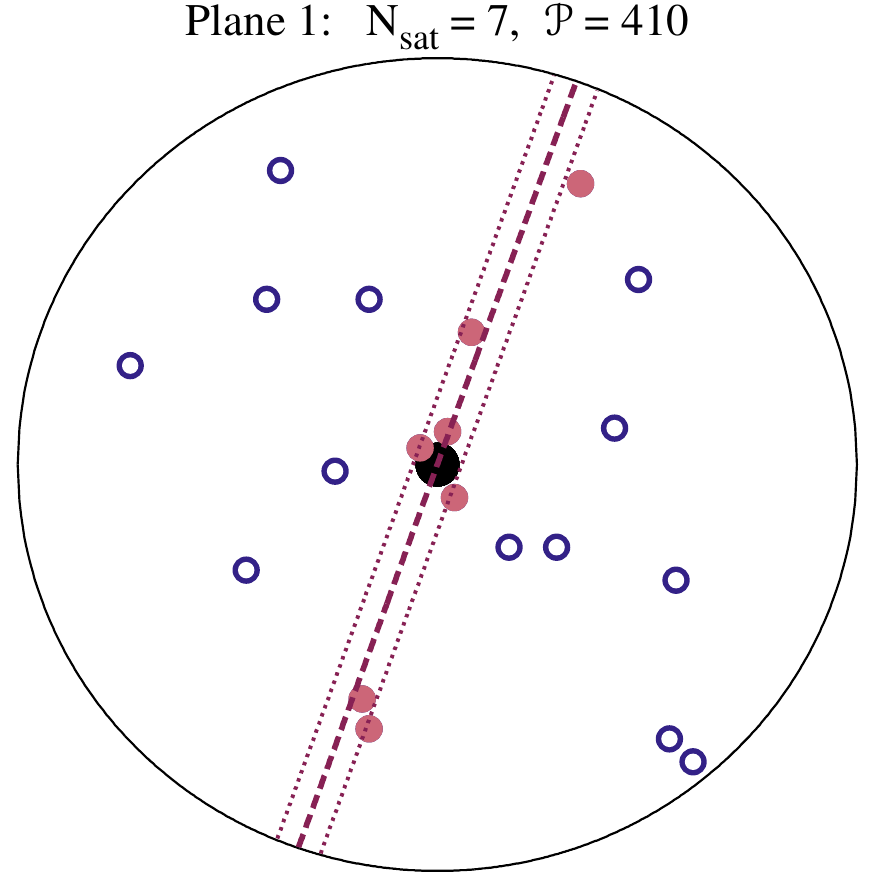} $\;\;\;\;$
     \includegraphics[width=0.3\linewidth,angle=0]{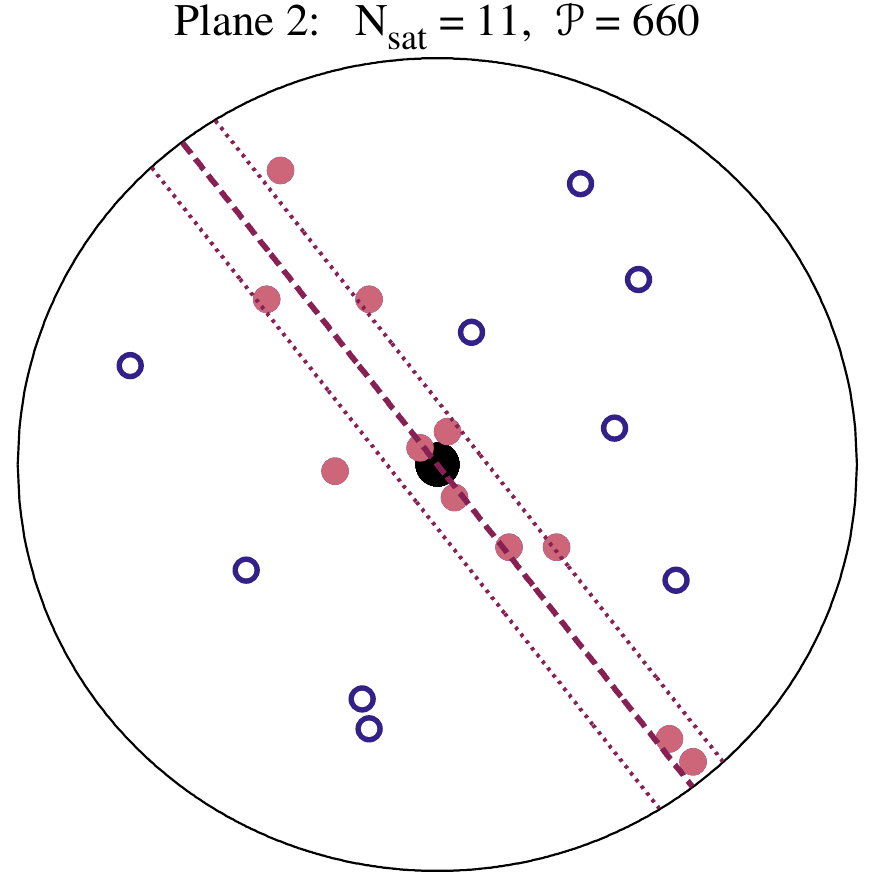} $\;\;\;\;$
     \includegraphics[width=0.3\linewidth,angle=0]{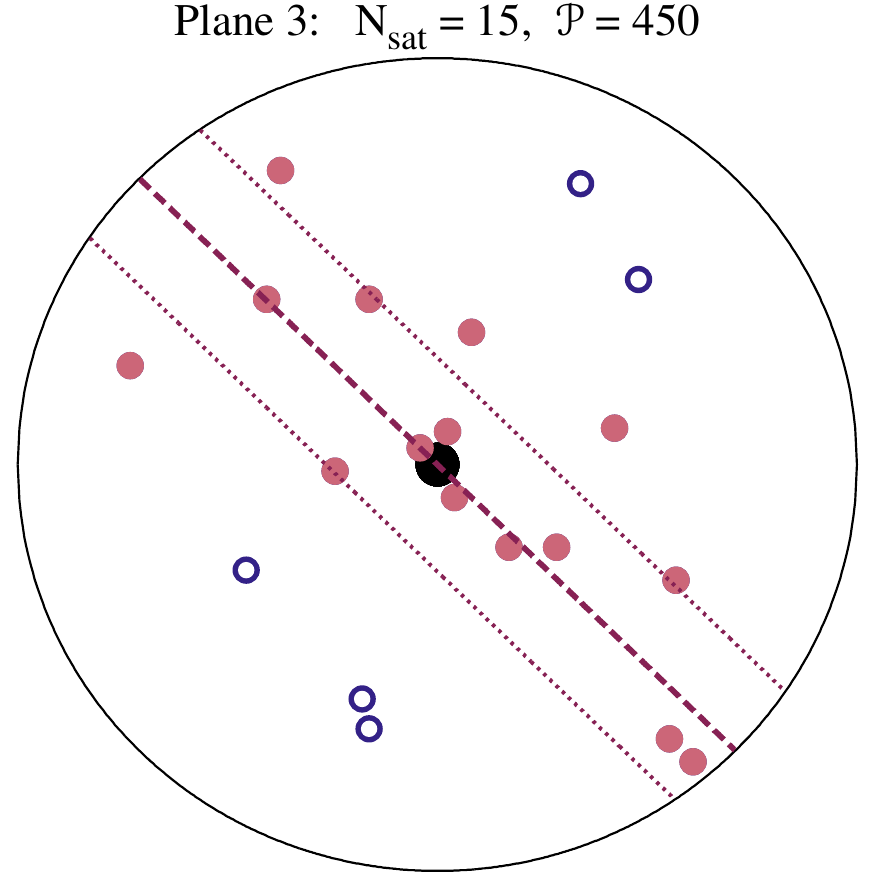}
     \caption{ Illustration of the procedure to identify planes of satellites. The panels show the same system, with the small open symbols corresponding to satellites identified around a central galaxy (large symbol in the centre). We take every subsample of any given number of satellites and compute the plane corresponding to that configuration. The panels shows three candidate configurations with $\Nsat=7$, $11$ and $15$ members (red filled symbols) out of a maximum of $\Nmax=20$ satellites. The dashed line corresponds to the best-fitting plane and the two dotted lines show the thickness of the plane, $\rPerp$ (see Eq. \ref{eq:plane_thickness}). To determine which configuration stands out the most, we compute the plane prominence, $\Prominence$. It specifies that the plane has a probability of one in $\Prominence$ to be a statistical fluctuation, e.g. plane 2 corresponds to one chance in 660 to be caused by a fluctuation. Out of the three, the rarest or most prominent plane is number 2 since it has the largest prominence. }
     \label{fig:illustrated_identification_method}
\end{figure*}

We select counterparts to the two massive members of the Local Group by identifying DM haloes with similar masses in the range $(1-3)\times10^{12}\Msun$, which is consistent with the mass of the MW and M31 halo \citep[e.g.][for a compilation of other measurements and discussions of systematic effects see \citealt{Coourteau2014,Wang2015}]{Fardal2013,Piffl2014,Cautun2014b,Cautun2014c,Veljanoski2014,Gonzalez2014}. This results in 2849 \MII{} haloes and in 63 \COCO{} haloes in the required mass range. Compared to the previous studies that analysed the incidence of the M31 plane of satellites \citep[e.g.][]{Bahl2014,Pawlowski2014c}, we adopted a broader mass range to account for the large uncertainty in the mass measurements and also for possible systematic effects. We checked that the exact mass range used does not affect our final results. 


\section{Identifying planar configurations of satellites}
\label{sec:defining_planes}

In this section we introduce an objective method to identify, for each halo, the rarest plane of satellites, both spatially and kinematically. The method works by identifying the subsample of satellites that is the least likely to be obtained by chance. This is motivated by recent observations that have shown that only a subset of the satellite galaxies are potentially distributed along a plane. For example, out of the 27 M31 satellites in the \pandas{} footprint, a significant plane is found for 15 of them (\ibata{}), while the entire population is no more planar than would be expected for an isotropic distribution of equal size \citep{Conn2013}. \citet{Tully2015} found that the Centaurus A Group shows evidence for two planes of galaxies, which, between them, contain 27 out of the 29 members with known distances. Even for the MW, while the 11 classical satellites are found on a plane, only 8 have orbital poles in a narrow angle indicating a possibly long lived planar structure \citep{Pawlowski2013b}. 

Previous methods of identifying planar configurations have been based on first examining the data and only in a second step defining an approach for selecting planes, resulting in a selection method that is both subjective and \textit{a posteriori}. For example, the 15 members of the M31 plane have been found by noticing that, when increasing the number of satellites associated with a plane, the best-fitting plane hardly changes and that the thickness of the plane increases only slowly. If one considers more than 15 satellites, this then leads to a larger change in the thickness of the best-fitting plane and in its direction (\ibata{}). Applying such a method to a large sample of systems is undesirable, since it implies choosing at least two threshold parameters: the maximum allowed changes in the thickness of a plane and in its direction when adding an extra valid plane member. There are no \textit{a priori} well-motivated values for those threshold parameters. In addition, the two thresholds should likely depend on the radial distribution of satellites, since more radially concentrated distributions will likely have thinner planes. While other methods of identifying planar distributions have been proposed \citep[e.g.][]{Conn2013,Gillet2015}, they all involve one or more subjective parameters. In contrast, the method we present here does not involve any such parameters and, in addition, it naturally takes into account the radial distribution of satellites in each system.

\subsection{Spatial planes}
\label{subsec:defining_spatial_planes}
We start by introducing a method for identifying the most prominent \planeOne{} plane. When comparing planes that contain the same number of satellites, the one that stands out the most is the thinnest plane. Difficulties arise when we have to compare planes with two different numbers of members, since it is not trivial to decide which one stands out more. In a nutshell, we identify all possible subsamples of satellites, out of a total sample of $\Nmax$ satellites, and, of those, we select the one configuration that is the least likely to be caused by statistical fluctuations. This is illustrated in \reffig{fig:illustrated_identification_method}.

In a first step, we identify the satellite subsets that are of interest for our study. We do so using the approach detailed in \refappendix{appendix:select_subset}. For each such subsample, we find the best-fitting plane, which is the plane that minimizes the root-mean-square distance of the satellites from it. For this, we define the plane thickness, $\rPerp$, as
\begin{equation}
    \rPerp = \sqrt{\frac{ \sum_{i=1}^{\Nsat} \left( \mathbf{n_{\rm plane}} \cdot \mathbf{x_i} \right)^2 } {\Nsat}}
    \label{eq:plane_thickness} \;,
\end{equation}
where $\Nsat$ is the number of satellites in the subsample and $\mathbf{n_{\rm plane}}$ denotes the normal to the plane. With $\mathbf{x_i}$ we denote the position of each satellite in a coordinate system whose origin is the central host galaxy. The plane thickness, $\rPerp$, is in fact the dispersion in the distance of the satellites from a plane that goes through the central galaxy. The best-fitting plane is the one that minimizes $\rPerp$. The normal to this plane, $\mathbf{n_{\rm plane}}$, is given by the eigenvector corresponding to the lowest eigenvalue of the inertia tensor of its members. 

Each resulting plane is characterized in terms of its prominence, $\Prominence$, such that, the larger the prominence, the least likely it is that the plane is due to a chance alignment. For example, for plane $i$ that has $N_{\rm sat;\;i}$ members and a thickness, $r_{\perp;\;i}$, the \pOne{} prominence is defined as:
\begin{equation}
    \Prominence_{\rm spatial}^{\rm plane\; i} = \frac{1}{ p\left( \le r_{\perp;\;i} \;| \;N_{\rm sat;\;i} \right) }
    \label{eq:p_spatial_one_plane} \;
\end{equation}
where the denominator gives the probability of obtaining by chance a configuration of $N_{\rm sat;\;i}$ satellites that is thinner than $r_{\perp;\;i}$. This probability is computed using $10^5$ isotropic distributions of satellites as outlined in \refappendix{appendix:probability}. Since the radial distribution of satellites has a strong effect on the thickness of the resulting planes, we generate each isotropic realization to have the exact same radial distribution as the system under study.

Now it is only natural to characterize the most prominent, or rarest, plane as the one that is the least likely to be obtained by chance. Using our notations, this can be written formally as:
\begin{align}
    \Pspatial & = \max_\textrm{all planes i} \left[\; \Prominence_{\rm spatial}^{\rm plane\; i} \;\right]  \label{eq:spatial_planes} 
    \;,
\end{align}
which says that the rarest \planeOne{} plane is the one that has the largest prominence. 
It is important to note that, within this approach, every halo contains a rarest plane. Determining if this rarest plane is statistically significant is a separate question that we will address in \refsec{sec:M31_planes}.

\subsection{Spatial and 2D-kinematic planes}
\label{subsec:defining_spatial_and_corotational_planes}
The observational data for the MW and M31 satellites contains both positions and velocity information for these objects. It is natural to try to incorporate this additional velocity information into the detection of planar configurations of satellites. The M31 satellites have only radial velocity measurements, so the full 3D velocities are unknown. But since the M31 plane of satellites is almost parallel to the line of sight, the radial velocities can be used to estimate the sense of rotation of each satellite with respect to the best-fitting plane. In the following, we describe how to select \pTwo{} planes, which are at the same time spatially thin and have a large number of members that share the same sense of rotation.

Before continuing, it is important to discuss some potentially misleading nomenclature used in previous studies. Satellites sharing the same sense of rotation have been referred to as corotating satellites \citep[e.g. \ibata{},][]{Bahl2014}. This nomenclature is confusing since corotation is normally used to denote a rotation around a common axis. Thus, two satellites corotate if their orbital poles are very close together. In the absence of 3D velocities, we only know that, when projected on the best-fitting plane, 13 out of the 15 satellites rotate in the same sense, either clockwise or counter-clockwise.

In addition to the steps described in \refsec{subsec:defining_spatial_planes}, for each satellite subset we also determine the number of members that share the same sense of rotation relative to the best-fitting plane. To determine the direction of rotation of each member, we take the scalar product between the plane normal and the orbital momentum of the satellite. A positive scalar product corresponds to clockwise rotation, while a negative one corresponds to counter-clockwise rotation. The number of satellites sharing the same sense of rotation, $\Ncor$, is the maximum of the number of objects rotating clockwise and those rotating counter-clockwise. Following this step, we assign to each plane a \pFour{} prominence, $\Prominence_{\rm 2D-kin}$, which is defined as:
\begin{equation}
    \Prominence_{\rm 2D-kin}^{\rm plane\; i} = \frac{1} { p\left( \ge N_{\rm \senseRot{};\;i} \;|\; N_{\rm sat;\;i} \right) }
    \label{eq:p_corotation_one_plane} \;,
\end{equation}
where the denominator gives the probability of obtaining by chance a configuration of $N_{\rm sat;\;i} $ satellites in which at least $N_{\rm \senseRot{};\;i}$ of them share the same sense of rotation. The procedure for computing this probability is detailed in \refappendix{appendix:probability}.

We define the rarest \pTwo{} planes as the one whose spatial and 2D kinematical distribution are the least consistent with a statistical fluctuation. Thus,
\begin{equation}
    \Pcombined = \max_\textrm{all planes i} \left[\; \Prominence_{\rm spatial}^{\rm plane\; i} \times  \Prominence_{\rm 2D-kin}^{\rm plane\; i}  \;\right]
    \label{eq:spatial_corotation_planes} \;,
\end{equation}
that is the plane that maximizes the product of the \pOne{} and the \pFour{} prominences.

\subsection{Spatial and 3D-kinematic planes}
\label{subsec:defining_spatial_and_orbital_planes}
In the case of the MW, the 3D velocities of the 11 classical satellites are known. This suggests that for the MW system one can identify planes that are both spatially thin and show a large degree of coherent 3D kinematics. For this, one needs to construct a cost function that rewards systems in which most satellites have orbital poles close together and penalizes systems in which the orbital poles are isotropically distributed. For example, to study long lived planar configurations, the cost function would preferentially reward systems in which the orbital momentum of its members is close to parallel or anti-parallel to the normal to the best-fitting plane. For this work, we employ a variant of the cost function suggested by \citet{Pawlowski2013b}, since this one has been used in other studies that claim a tension between the MW satellite plane and \lcdm{} predictions \citep[e.g.][]{Pawlowski2014c}. That function has been proposed after examining the orbital data of the MW satellites, and as such it is an \textit{a posteriori} definition of the rotation characteristics of the Galactic satellite distribution. It may be that other satellite planes in the Universe have different orbital structures, in which case that cost function may not be optimally suited for characterizing their kinematical structure. 

\begin{table*}
    \begin{minipage}{\textwidth}
    
    \begin{center}
    \renewcommand{\arraystretch}{1.5}
    \caption{ The characteristics of the rarest planar configuration of satellites around the M31 and the MW.} 
    \label{tab:plane_properties}
    \begin{tabular}{ L{0.2cm} L{3.3cm} ll L{1.05cm}L{1.05cm} L{1.2cm} L{2.2cm} L{1.5cm} L{1.5cm}}
        \hline
        ID & Plane type    &  $\Nsat$  &  $\Ncor$  &  $\rPerp$ $({\rm kpc})$  &  $\rAlong$ $({\rm kpc})$ & $\Dorbit$ (degrees)   & $\Prarest$  & \multicolumn{2}{p{3.5cm}}{fraction of systems with more prominent planes $(\%)$} \\[-0.1cm]
         & & & &  &  &  & & \lcdm{} & isotropic \\[-0.05cm]
        \hline\hline
        \multicolumn{10}{p{.88\textwidth}}{}\\[-0.3cm]
        \multicolumn{10}{p{.88\textwidth}}{\centering M31 plane of satellites} \\
        \hline
        (1) & \plOne{}     &   14 &   11$^a$  &  $10.3^{+0.7}_{-0.6}$ &   $220^{+23}_{-24}$  &   - &  $1.0^{+1.1}_{-0.5} \times 10^3$ &  $12^{+6}_{-4}$      &  $1.2^{+1.0}_{-0.6}$    \\
        (2) & \plTwo{}     &   15 &   12$^a$ &  $12.5^{+0.7}_{-0.5}$ &   $214^{+22}_{-23}$  &   - &  $3.4^{+2.8}_{-1.6} \times 10^4$ &  $8.8^{+2.8}_{-1.8}$ &  $0.34^{+0.38}_{-0.14}$ \\
        \hline\hline
        \multicolumn{10}{p{.88\textwidth}}{}\\[-0.3cm]
        \multicolumn{10}{p{.88\textwidth}}{\centering MW plane of satellites} \\
        \hline
        (3) & \plOne{}      &   11 &     8$^b$ &  $20.7^{+0.6}_{-0.6}$ &   $129^{+4}_{-4}$  &  $63^{+6}_{-3}$  &  $2.6^{+0.4}_{-0.4} \times 10^2$ &  $12^{+1}_{-1}$      &  $2.9^{+0.4}_{-0.4}$    \\
        (4) & \plTwo{}     &   11 &     8$^b$ &  $20.7^{+0.6}_{-0.6}$ &   $129^{+4}_{-4}$  &  $63^{+6}_{-3}$  &  $1.5^{+2.5}_{-0.7} \times 10^3$ &  $10^{+5}_{-5}$ &  $2.0^{+1.5}_{-1.1}$    \\
        (5) & \plThree{}   &   11 &     8$^b$ &  $20.7^{+0.6}_{-0.6}$ &   $129^{+4}_{-4}$  &  $63^{+6}_{-3}$  &  $8.3^{+32}_{-6.0} \times 10^3$ &  $5.0^{+4.1}_{-2.7}$ &  $0.48^{+0.87}_{-0.39}$ \\ 
        \hline
    \end{tabular}
    \end{center}
    
    \small  \smallskip
    The M31 plane was found using 27 satellites within the \pandas{} footprint. The MW plane was found using the brightest 11 classical satellites and a Galactic obscuration zone of $33\%$ of the sky. The table columns give: a plane ID for easy reference, plane selection method, the number of satellites in the plane ($\Nsat$) and how many of those share the same sense of rotation with respect to the best-fitting plane($\Ncor$), the plane thickness ($\rPerp$) and radial extent ($\rAlong$), the orbital pole dispersion for the plane members ($\Dorbit$), the prominence of the plane ($\Prarest$), the fraction of systems with more prominent planes, i.e. higher $\Prarest$, for \lcdm{} and isotropic satellite distributions.
    
    \smallskip   \smallskip
    $^{a}$ The sense of rotation of And XXVI is highly uncertain since it depends on the object's radial distance from the MW. And XXVI is more likely to counter-rotate ($55\%$ probability) than to rotate in the same sense as the majority of the satellites in the plane ($45\%$ probability). This is in contradiction with the \ibata{} results, who claimed that And XXVI is one of the 13 plane members that share the same sense of rotation. We suspect that \ibata{} calculated the sense of rotation using the distance corresponding to the peak of the radial distance PDF, which indeed would result in the claimed result. The radial distance PDF of And XXVI is highly asymmetrical, so the position of the PDF peak does not necessarily characterize the most likely outcome.
    
    $^{b}$ Due to large proper motion errors, there are uncertainties in the sense of rotation of: Sextans, Carina, Leo I and Leo II. This mean that the number of satellites sharing the same sense of rotation is 6, 7, 8 or 9 with a probability of $1$, $16$, $50$ and $33\%$, respectively.
    \end{minipage}
\end{table*}

To compute the amount of kinematical information, we proceed as follows. For each of the satellite subsets used in \refsec{subsec:defining_spatial_planes}, we determine the dispersion in the orbital poles, i.e. directions of the orbital momenta, of its members as
\begin{equation}
    \Dorbit = \sqrt{\frac{ \sum_{i=1}^{\Nsat} \arccos^2\left( \mathbf{\overline{n}_{\rm orbit}} \cdot \mathbf{n_{\rm orbit; \; i}} \right) } {\Nsat}}
    \label{eq:orbital_dispersion}
\end{equation}
where $\mathbf{n_{\rm orbit; \; i}}$ denotes the orbital momentum direction of each member of the plane and $\mathbf{\overline{n}_{\rm orbit}}$ the mean orbital pole of all the $\Nsat$ members found in the plane. Compared to our approach, \citet{Pawlowski2014c} applied \eq{eq:orbital_dispersion} to only 8 out of the 11 satellites found in the MW satellite plane. This was motivated by the observation that only 8 out of the 11 members show close orbital poles\footnote{Incorporating such a posteriori considerations incurs the danger of designing tests that are specifically matched to one particular system and that may not be characteristic of the larger population.}. In principle, we could follow a similar approach and take a subset of the plane members that shows the most concentrated orbital poles. This would amount to taking a subset of a subset, since our planes are already subsets of satellites from a larger sample. We prefer not to do so since it would add an additional layer of complexity to this method and also a significantly higher computational cost.

After applying \eq{eq:orbital_dispersion} to each plane, we define the \pFive{} prominence, $\Prominence_{\rm 3D-kin}$, of plane $i$ as
\begin{equation}
    \Prominence_{\rm 3D-kin}^{\rm plane\; i} = \frac{1}{ p\left( \le \Delta_{\rm std;\;i} \;| \;N_{\rm sat;\;i} \right) }
    \label{eq:p_orbit_one_plane} \;,
\end{equation}
where $\Delta_{\rm std;\;i}$ denotes the orbital dispersion of the plane. The denominator gives the probability of obtaining by chance a configuration of $N_{\rm sat;\;i}$ that have an orbital pole dispersion less than $\Delta_{\rm std;\;i}$. The procedure for computing this probability is given in \refappendix{appendix:probability}. 

Now we can define the prominence of the rarest \planeThree{} plane as
\begin{equation}
    \PcombinedMW = \max_\textrm{all planes i} \left[\; \Prominence_{\rm spatial}^{\rm plane\; i} \times  \Prominence_{\rm 3D-kin}^{\rm plane\; i}  \;\right]
    \label{eq:spatial_orbit_planes} \;,
\end{equation}
which is the plane of satellites whose spatial and 3D kinematic distribution is the least likely to be a statistical fluctuation.


\section{M31-like planes of satellites}
\label{sec:M31_planes}

Here we investigate the characteristics of the rarest planes of satellite galaxies as found in mock \pandas{}-like catalogues. The goal is to obtain a better understanding of the M31 plane of satellites and to compare it to the \lcdm{} predictions.

To create \pandas{}-like mocks, we use the host halo sample described in \refsec{sec:data}. For each of those hosts, we start by finding all the satellites with stellar masses larger than $2.8\times10^4\Msun$ \citep[as proposed by][]{Bahl2014} that are within a radial distance of up to $500\kpc$. To reproduce the observational geometry, we place the observer at a distance of $780\kpc$ from the centre of the host halo, which is equivalent to the MW-M31 distance \citep{Conn2012}. For each satellite identified earlier, we compute its sky coordinates, as seen by the observer. Out of all the satellites within the \pandas{} mask that are also more than $2.5^\circ$ from the host (\ibata{}), we keep only the 27 objects that have the largest stellar masses. If there are fewer than 27 satellites within the required geometry, we discard that host halo. For each host, we place the line-of-sight along three mutually perpendicular directions consisting of the simulation's x, y and z axes. Due to the highly asymmetrical \pandas{} volume, this will result in somewhat different satellite distributions, hence increasing the overall statistics. After applying this procedure, we end up with 7350 mock satellite systems in \MII{} and 180 in \COCO{}. 

\subsection{The M31 system}
In a first step, we apply our method to the actual \pandas{} observations of M31. We do not use the same plane identification method as \ibata{}, so it is important to check what it is that our approach identifies as the most prominent plane of the M31 system. To account for observational errors, we generate 1000 Monte Carlo realizations that sample the radial distance PDFs \citep[][Table 1]{Conn2012} and radial velocities \citep[][Table 5]{Collins2013} of the M31 satellites.

In the case of the M31 system, the observational data allow for the identification of \pOne{} and \pTwo{} planes. For each Monte Carlo realization of the M31 system, we identify the rarest plane. Due to the large radial distance errors, the rarest plane can vary between realizations. For example, the rarest \planeOne{} plane contains 14 members in $72\%$ of the cases and 13 members in $22\%$ of realizations. In the remaining $6\%$ of realizations it contains even fewer satellites. For simplicity, we take the rarest plane as the one that is identified as such in the largest number of realizations. The rarest planar configurations of the M31 system and its characteristics are shown in \reftab{tab:plane_properties}. 
 
We find that the rarest \planeOne{} plane consists of 14 satellites that are the same as the 15 members of the plane reported by \ibata{}, except And III. This is in agreement with the results of \ibata{}, who point out that choosing 13 or 14 satellites results in a higher spatial significance, i.e. lower probability of being obtained from an isotropic distribution, than for the full sample of 15. 
The \planeTwo{} plane found by our approach is the same as the one reported by \ibata{}, even though our plane identification method is different. \ibata{} reported the significance of the M31 plane as compared to an isotropic distribution, so it is possible that they inadvertently choose the parameters of their method (see \refsec{sec:defining_planes}) such that it maximizes the plane significance. If that was the case, then both plane finding methods are basically the same.


\subsection{The rarest M31-like planes}
\label{subsec:M31_rarest_planes}

\begin{figure}
     \centering
     \includegraphics[width=1.03\linewidth,angle=0]{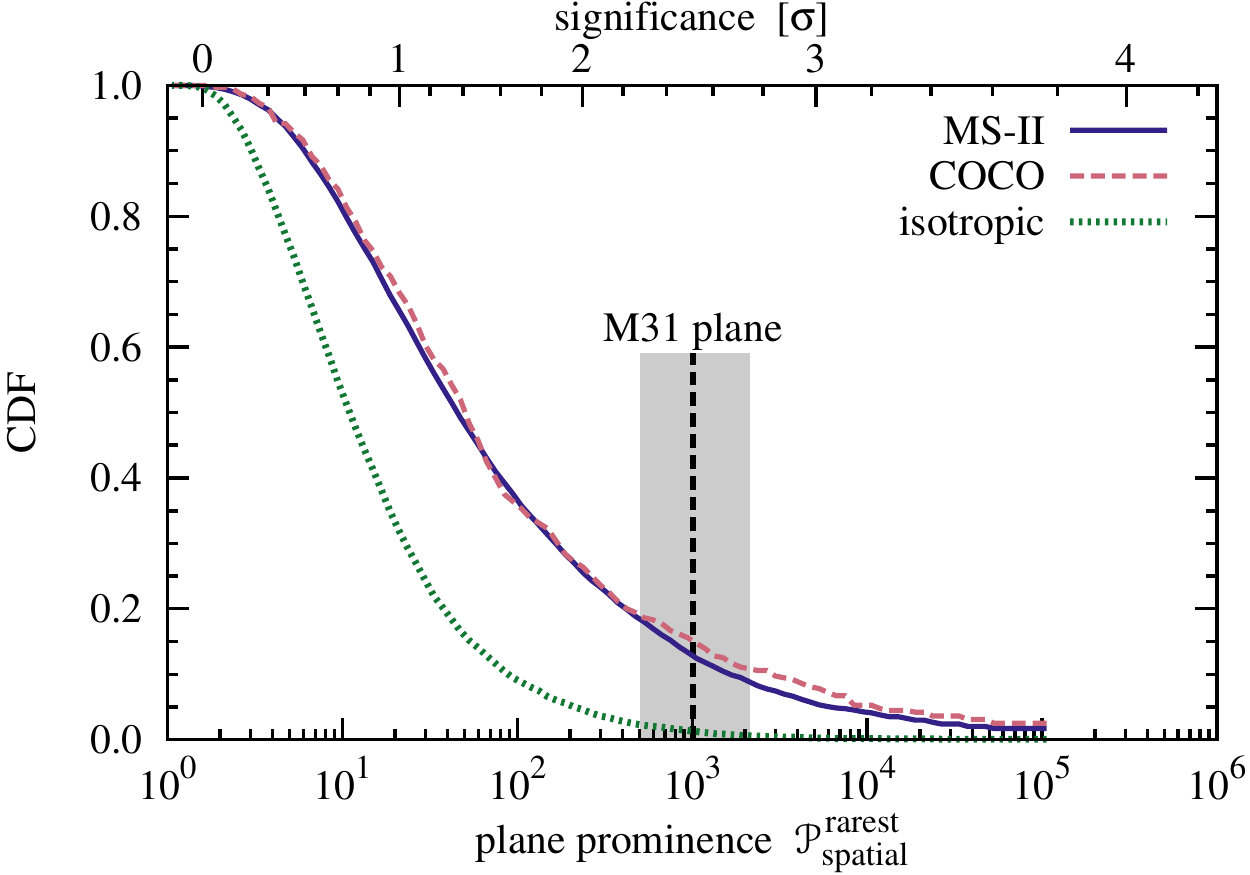} 
     \caption{ The CDF of the prominence, $\Pspatial{}$, of the rarest \planeOne{} plane of satellites for mock \pandas{} observations. The solid line gives the \MII{} results, while the dashed line shows results for \COCO{}, which has much higher resolution. The dotted curve gives the expectation for isotropic satellite distributions. The vertical dashed line and shaded region show the prominence and the $1\sigma$ error for M31's \planeOne{} plane of satellites. We find that $12^{+6}_{-4}\%$ of \lcdm{} haloes have a more prominent planar configuration than M31. The top x-axis shows the detection significance of each plane computed using the isotropic CDF (dotted curve) and accounts for the \emph{look-elsewhere effect}. }
     \label{fig:M31_spatial_plane_PDF}
     
     \includegraphics[width=1.03\linewidth,angle=0]{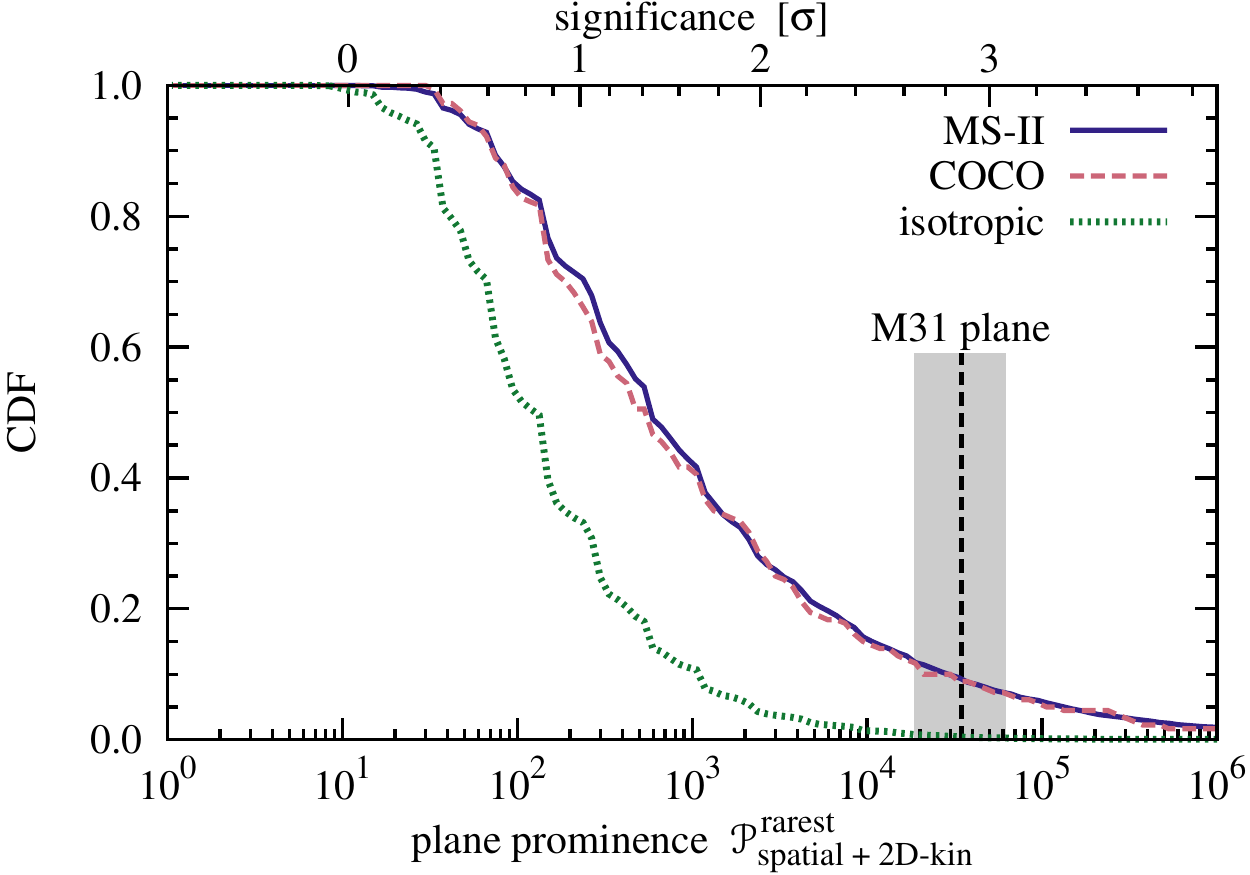}
     \caption[]{ As \reffig{fig:M31_spatial_plane_PDF}, but for the prominence, $\Pcombined{}$, of the rarest \planeTwo{} plane of satellites. This case corresponds to the M31 plane identified by \ibata{}. For this test, $8.8^{+2.8}_{-1.8}\%$ of \lcdm{} hosts have a more prominent plane than M31.}
     \label{fig:M31_combined_plane_PDF}
\end{figure} 

To better understand the M31 plane of satellites, we start by assessing the chance of obtaining more prominent planar configurations within \lcdm{}. This is shown in \reffig{fig:M31_spatial_plane_PDF}, where we plot the cumulative distribution function (CDF) of the prominence, $\Pspatial{}$, of the rarest \planeOne{} planes. There is a very good match between the \MII{} and \COCO{} haloes, which suggests that satellite planes found in \MII{} are not significantly affected by the limited resolution of the simulation. We find that most haloes have prominent planar configurations, for example 37, 12 and $4\%$ of haloes have planes with $\Pspatial\ge10^2$, $10^3$ and $10^4$ respectively. 

\MCn{The prominence is not simply the inverse of the probability for isotropic satellite distributions for reasons that will be discussed in \refsec{subsec:M31_look_elsewhere}.} As such, the figure also shows the rarest planes found in an isotropic distribution of satellites. To obtain these, for each of the \lcdm{} haloes, we generate an isotropic distribution by choosing random polar and azimuthal angles, while keeping the radial position of each satellite fixed. We then apply the same plane identification procedure to each isotropic satellite distribution. Unsurprisingly, we find a clear difference between the isotropic and \lcdm{} results, with the isotropic CDF shifted towards the left of the \lcdm{} CDF. This suggest that, compared to a uniform distribution, there is more structure in the distribution of \lcdm{} satellites, in agreement with previous studies \citep[e.g.][]{Libeskind2005,Wang2013,Pawlowski2014c}. 

The corresponding M31 plane, entry (1) from \reftab{tab:plane_properties}, is shown as the dashed vertical line. It has a prominence, $\Pspatial=1.0^{+1.1}_{-0.5} \times 10^3$, which means that for an isotropic distribution there is only a 1 in 1000 chance of obtaining a thinner plane with $14$ members. This result was found by computing the prominence of the rarest plane for each Monte Carlo realization of the M31 system. Following this, we quote the median value and the 1$\sigma$ interval, corresponding to the 16th to 84th percentiles. We find that the \planeOne{} plane of the M31 system is consistent with \lcdm{} expectations, since there is a $12^{+6}_{-4}\%$ chance of finding an even more prominent plane in \lcdm{}. In fact, the M31 \planeOne{} plane is consistent, at $2.5\sigma$, even with an isotropic distribution, since there are $1.2^{+1.0}_{-0.6}\%$ more extreme systems in this case (these results are summarized in \reftab{tab:plane_properties}).

\reffig{fig:M31_combined_plane_PDF} shows the prominence of the rarest \planeTwo{} planes. The general conclusions are the same as for the previous figure: we find a very good match between the \MII{} and the \COCO{} results; and \lcdm{} satellite distributions have more prominent planes than isotropic distributions. The bumpy aspect of the CDF curves for the \MII{} and the isotropic case reflects the discrete nature of the 2D-kinematical test, since the number of plane members sharing the same sense of rotation always takes integer values. For this test, the corresponding M31 plane, entry (2) from \reftab{tab:plane_properties}, which is the one identified by \ibata{}, is characterized by a prominence, $\Pcombined{}=3.4^{+2.8}_{-1.6} \times 10^4$. There are $8.8^{+2.8}_{-1.8}\%$ \lcdm{} haloes with more prominent planes suggesting that the M31 plane of satellites is in agreement with \lcdm{} predictions.

\subsection{The detection significance of a plane}
\label{subsec:M31_look_elsewhere}
We now discuss the detection significance of a plane of satellites, i.e the probability that a plane is due to a statistical fluctuation. For this, we need to take into account the \emph{``look-elsewhere" effect}. This is a phenomenon in statistics where an apparently  statistically significant observation may have actually arisen by chance because of the large size of the parameter space to be searched. It represents an important effect for cases where one does not have an \emph{a priori} model or prediction to where the signal should appear, and, hence, when one needs to search for a signal in a large range. In such cases, the significance calculation must take into account that a high statistical fluctuation anywhere in that range could also be considered as a signal \citep[e.g. see][for a more rigorous discussion]{Gross2010}. The effect is particularly relevant in particle physics, and, in general, in any field in which one searches for uncommon events. 

The look-elsewhere effect is important since we do not know \emph{a priori} what is the number of satellites that we expect to find in a plane. In the case of \pandas{}-like observations, there are 27 satellites in each system, which means that the most prominent plane can have anywhere between 3 to 27 members. To estimate the significance of a plane, we need to compute the probability that a statistical fluctuation generating such a prominent plane appeared for any combination of $3,4,\ldots,27$ satellites \textemdash{} this is the \emph{marginalized probability}. This is different from the \emph{conditional probability} that has the number of satellites chosen as a prior and whose inverse gives the prominence of a plane (e.g. Eq. \ref{eq:p_spatial_one_plane}). The marginalized probability is the isotropic CDF shown in \reffigS{fig:M31_spatial_plane_PDF}{fig:M31_combined_plane_PDF}. In other words, the probability that a plane is a statistical fluctuation is given by the fraction of isotropic realizations that have a more prominent plane. Using this, we compute the significance of each plane, which is shown as the top x-axis of \reffigS{fig:M31_spatial_plane_PDF}{fig:M31_combined_plane_PDF}, with the tick marks spaced at $0.2\sigma$ intervals. The significance is expressed in multiples of the standard deviation, $\sigma$, of a normal distribution. Note that the one-to-one map between the prominence (bottom x-axis) and the significance level (top x-axis) differs between the two figures. 

\begin{figure}
     \centering
     \includegraphics[width=1.\linewidth,angle=0]{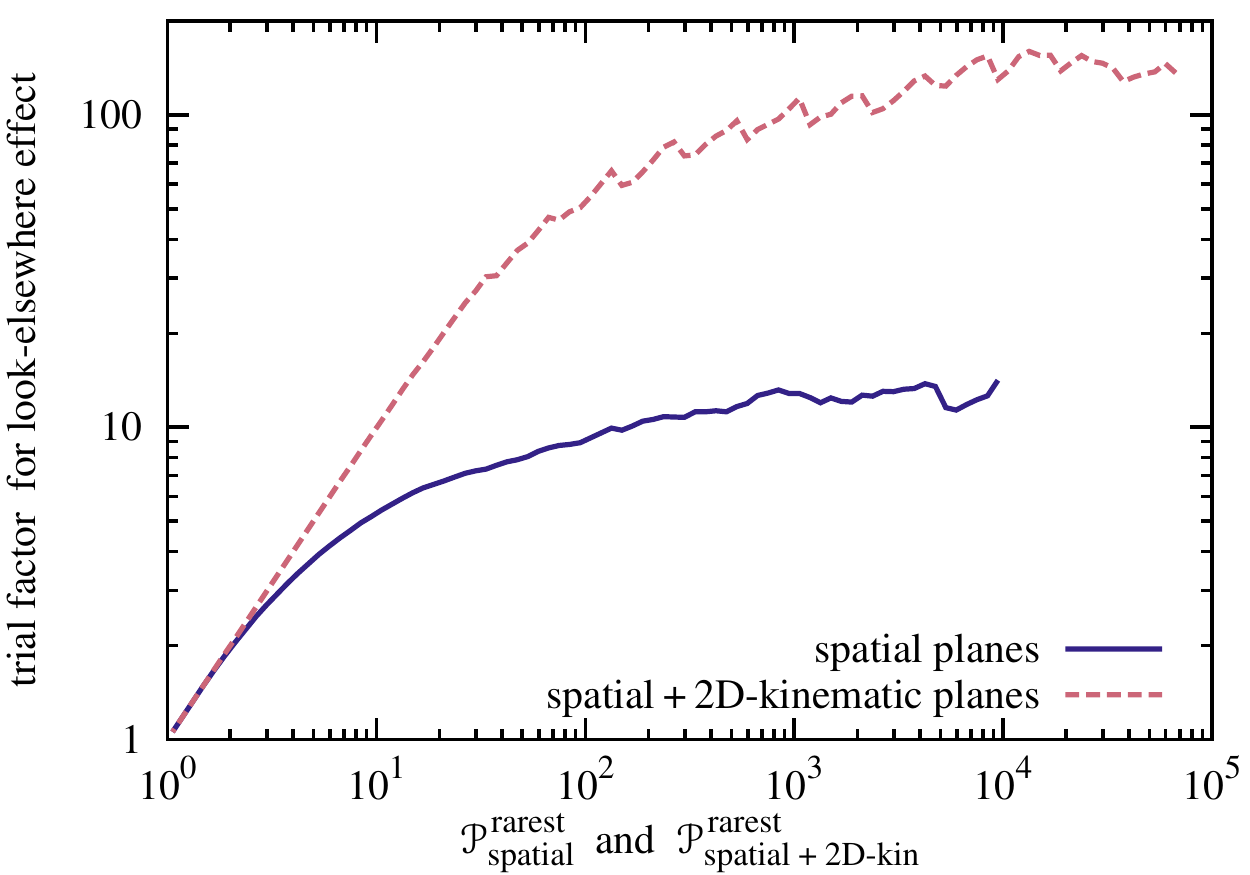}
     \caption{ The trial factor for the \emph{look-elsewhere effect} as a function of the prominence of the rarest plane. The ratio, ${\rm trial\;factor} / \Prarest$, gives the marginalized probability of obtaining by chance (i.e. in an isotropic distribution) a plane with prominence, $\Prarest$. The regions of interest are roughly $\Pspatial\ge10$ and $\Pcombined{}\ge100$ corresponding to the intervals in \reffigS{fig:M31_spatial_plane_PDF}{fig:M31_combined_plane_PDF} where the CDF is different from unity. }
     \label{fig:M31_trial_factors}
\end{figure} 

Some previous studies have incorrectly referred to the conditional probability as the detection significance of planes of satellites \citep[e.g.][\ibata{}]{Kroupa2005}. To better emphasize the difference between the two, we introduce the concept of \emph{trial factor} \citep[e.g. see][]{Gross2010}. This is the ratio between the marginalized and the conditional probability to obtain a statistical fluctuation with prominence, $\Prominence$. The former corresponds to the actual detection significance and is given by the CDF of an isotropic distribution. The latter corresponds to $\Prominence^{-1}$ since this is the definition of a plane's prominence (e.g. Eq. \ref{eq:p_spatial_one_plane}). Thus,
\begin{align}
    {\rm trial\;factor} & = \frac{{\rm CDF}_{\rm isotropic}(\ge \Prarest)}{\left(\Prarest\right)^{-1}} \nonumber \\
                        & = \Prarest \; {\rm CDF}_{\rm isotropic}(\ge \Prarest) \;.
\end{align}
For example, the M31 \planeTwo{} plane has a trial factor of 115 (see entry (2) in \reftab{tab:plane_properties} for numerical values). Thus, the chance of it being a statistical fluctuation is 115 times higher than naively expected if one considers only random planes with 15 members. Inevitably, this means that \ibata{} have overestimated the detection significance of the M31 plane by more than two orders of magnitude. For an isotropic distribution there is a $0.34\%$ probability of obtaining a more prominent plane, and, hence, the M31 plane corresponds to a $2.9\sigma$ detection.

In \reffig{fig:M31_trial_factors} we show the trial factors for \pOne{} and \pTwo{} planes. The regions of interest are roughly $\Pspatial\ge10$ and $\Pcombined{}\ge100$ corresponding to the intervals in \reffigS{fig:M31_spatial_plane_PDF}{fig:M31_combined_plane_PDF} where the isotropic CDF is different from unity. In those intervals, the trial factors increase only slowly with the plane prominence, so, to a first approximation, the two plane types have a trial factor of ${\sim}10$ and ${\sim}100$, respectively. The \pTwo{} planes have a higher trial factor due to the larger range used to search for the most prominent plane, since, on top of the spatial distribution, also the 2D kinematics are considered.

\subsection{The characteristics of rare planes}
\label{subsec:M31_plane_characteristics}
\begin{figure}
     \centering
     \includegraphics[width=1.\linewidth,angle=0]{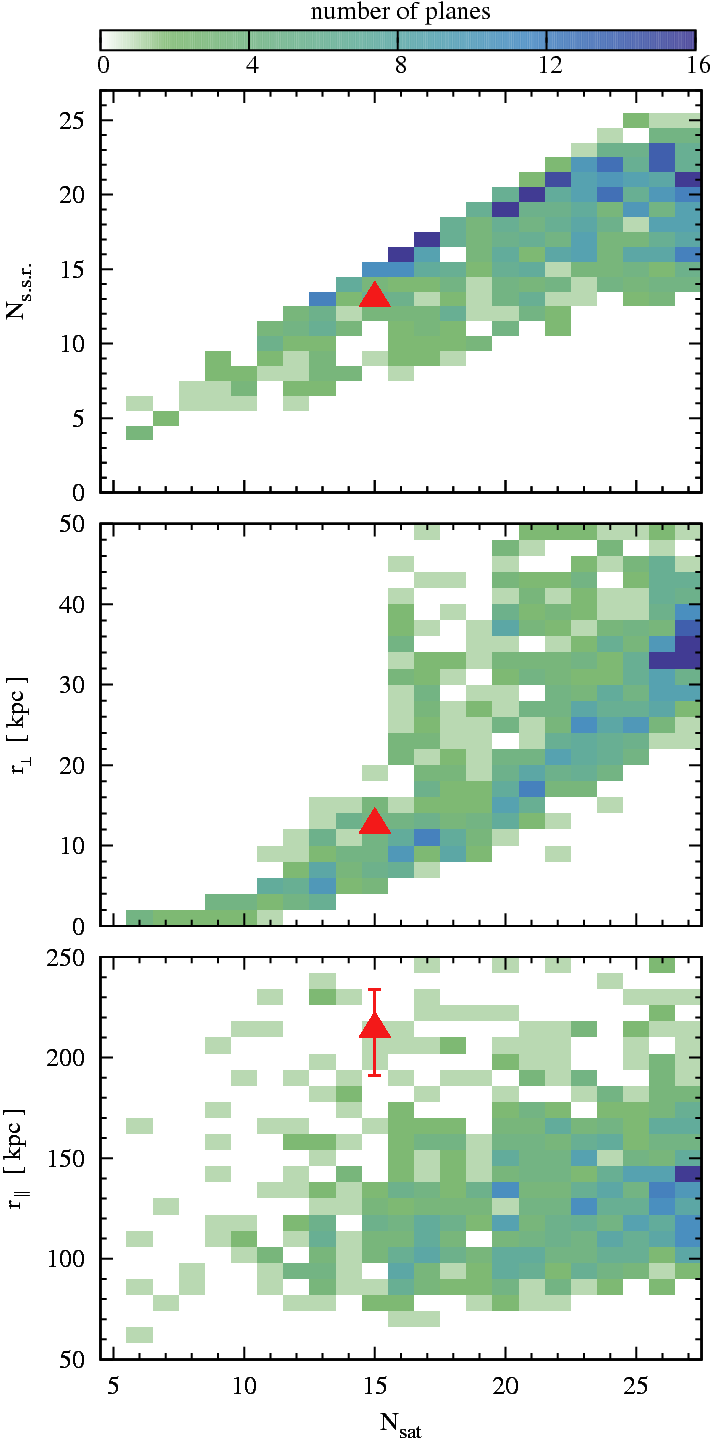}
     \caption{ The characteristics of the rarest \pTwo{} planes of satellites that are at least as prominent as the M31 plane, i.e. $\Pcombined{} \ge 1.8\times10^{4}$. There are 852 such systems in \MII{}. The grid cells are coloured according to the number of systems with those properties. The three panels show the number of satellites sharing the same sense of rotation, $\Ncor$ (top), the thickness, $\rPerp$ (centre), and the radial extent, $\rAlong$ (bottom), of the best-fitting plane, as a function of the number of satellites in the plane, $\Nsat$. The large triangle shows the properties of the M31 plane.}
     \label{fig:M31_2d_histogram_Nsat}
\end{figure} 

In \reffig{fig:M31_2d_histogram_Nsat} we show the characteristics of the rarest \planeTwo{} planes that are at least as prominent as the M31 plane. We describe the planes in terms of the number of members sharing the same sense of rotation, the plane thickness and the radial extent of the plane, $\rAlong$. This latter property characterizes the dispersion of the satellites within the plane and it is calculated as the mean sum of the squares of the distance projected on to the best-fitting plane. We choose these plane characteristics to be consistent with previous studies that investigated the incidence of the MW and M31 plane of satellites in term of these properties \citep[e.g.][]{Bahl2014,Ibata2014b}.

\reffig{fig:M31_2d_histogram_Nsat} shows that there is considerable variation among the properties of the most prominent planes, suggesting that each plane is different. For example, the number of members sharing the same sense of rotation, $\Ncor{}$, can take values between $\Nsat/2$ to $\Nsat$. The top panel of the figure shows that the planes have $\Ncor{}$ values spanning the full allowed range, although there is a higher preference for $\Ncor{}\simeq\Nsat$, since that will result in a higher prominence. The behaviour in the middle panel is governed by two requirements. Firstly, to be prominent, planes with a small number of members need to be very thin since such structures cannot have a high \pFour{} prominence, which explains the distribution seen in the left-half of the panel. Secondly, once the number of members is high enough, ${\sim}16$ in this case, the \pFour{} prominence can be by itself very large, so that such planes do not necessarily need to be very thin. This explains the large scatter in the $\rPerp{}$ values seen in the right-half of the middle panel. And lastly, the prominence of a plane does not depend on $\rAlong{}$, which explains the large scatter in $\rAlong{}$ values seen in the bottom panel of the figure.

\reffig{fig:M31_2d_histogram_Nsat} also shows the properties of the M31 plane of satellites whose position is marked with a large triangle. The M31 plane is within the scatter expected for \lcdm{} planes, \MCn{although it does stand out as having an unusually large radial extent, $\rAlong$}.

\subsection{The incidence of rare planes}
\label{subsec:M31_plane_incidence}
For each halo, we study the incidence of the rarest \pTwo{} plane among the distribution of satellites of all other \lcdm{} haloes. This is motivated by the studies of \citet{Ibata2014b} and \citet{Pawlowski2014c} that interpreted the low incidence of the M31 plane as evidence for an inconsistency between observed planes of satellites and \lcdm{} predictions.

\begin{table*}
    \begin{minipage}{\textwidth}
    
    \begin{center}
    \renewcommand{\arraystretch}{1.5}
    \caption{ The incidence of the M31 and the MW plane of satellite galaxies.} 
    \label{tab:plane_incidence}
    \begin{tabular}{ L{0.3cm} L{.5cm} L{5cm} L{5cm} L{3.5cm} }
        \hline
        ID & Host    &  $\fractLCDM$ &  $\fractLCDMorbit$  & fraction of \lcdm{} systems with lower frequencies $(\%)$ \\[-0.1cm]
        \hline\hline
        (2) & M31 &   $6.8^{+7.2}_{-2.7}\times10^{-4}$  &   - &  $5.1^{+4.5}_{-0.9}$  \\
        (4) & MW  &   $4.6^{+2.8}_{-2.5}\times10^{-3}$  &   - &  $18^{+6}_{-8}$  \\
        (5) & MW  &   -   &   $1.5^{+1.9}_{-1.2}\times10^{-3}$  &  $11^{+6}_{-7}$  \\ 
        \hline
    \end{tabular}
    \end{center}
    
    \small  \smallskip
    The table columns give: the plane ID from \reftab{tab:plane_properties}, the central galaxy (M31 or MW), the incidence, $\fractLCDM$ and $\fractLCDMorbit$, of similar planes among the \MII{} haloes, and the fraction of \lcdm{} haloes that have planes of satellites with even lower frequencies.
    \end{minipage}
\end{table*}

\begin{figure}
     \centering
     \includegraphics[width=1.\linewidth,angle=0]{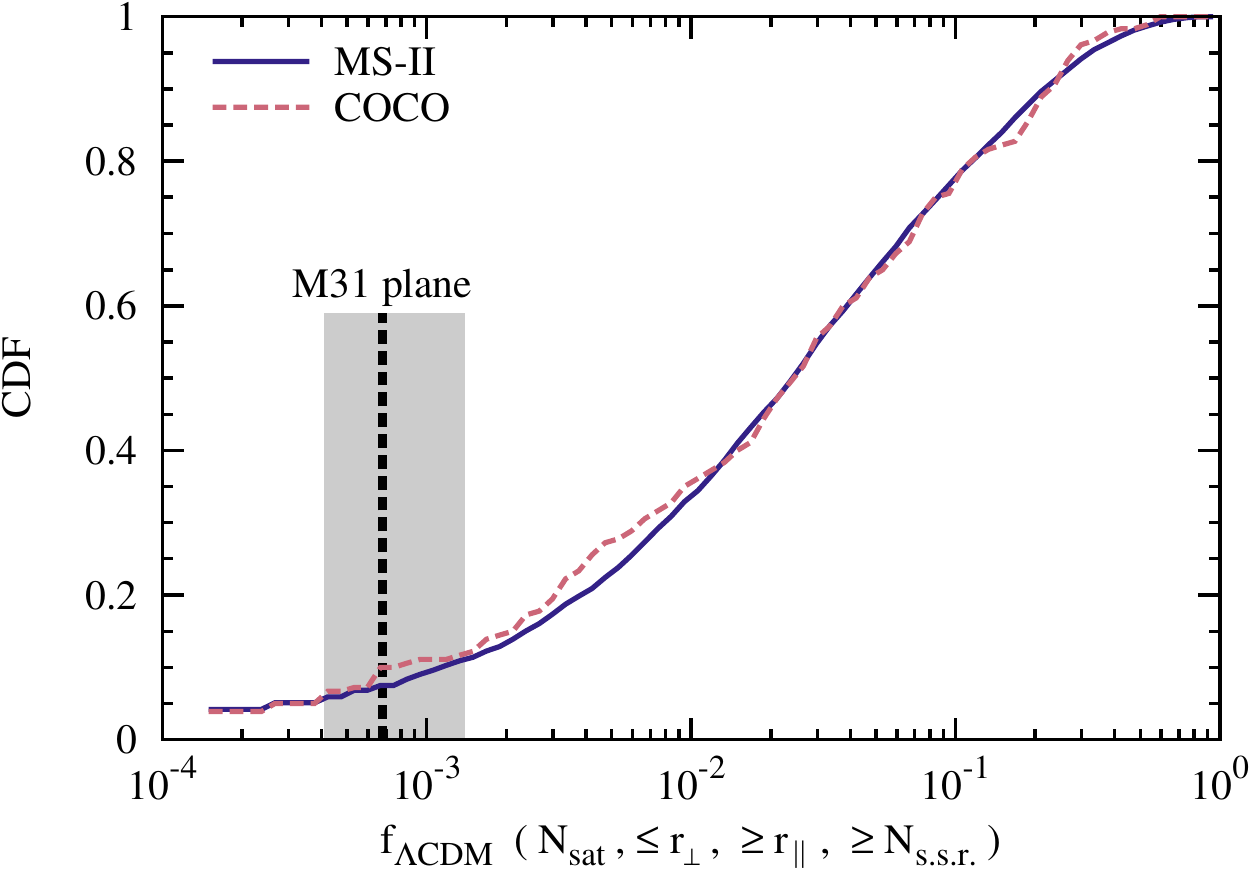}
     \caption[]{ The CDF of the incidence, $\fractLCDM$, of similar planes to the rarest one in \lcdm{}. For each \lcdm{} halo, we take the most prominent plane and find its frequency among all other \lcdm{} haloes (see the text for more details). Most planes have a low incidence (half of systems have a frequency of $0.02$ or lower) indicating that each \lcdm{} halo has a different planar configuration. The vertical dashed line and the grey area show the incidence and the $1\sigma$ error for the M31 plane. We find that $5.1^{+4.5}_{-0.9}\%$ of \lcdm{} systems have a lower frequency than the M31 plane. Thus, the low incidence of the M31 plane is not in tension with \lcdm{}, as claimed by \citet{Pawlowski2014c}, but instead is consistent with \lcdm{} expectations. }
     \label{fig:M31_fraction_haloes_with_similar_planes}
\end{figure} 

\begin{figure}
     \centering
     \includegraphics[width=1.\linewidth,angle=0]{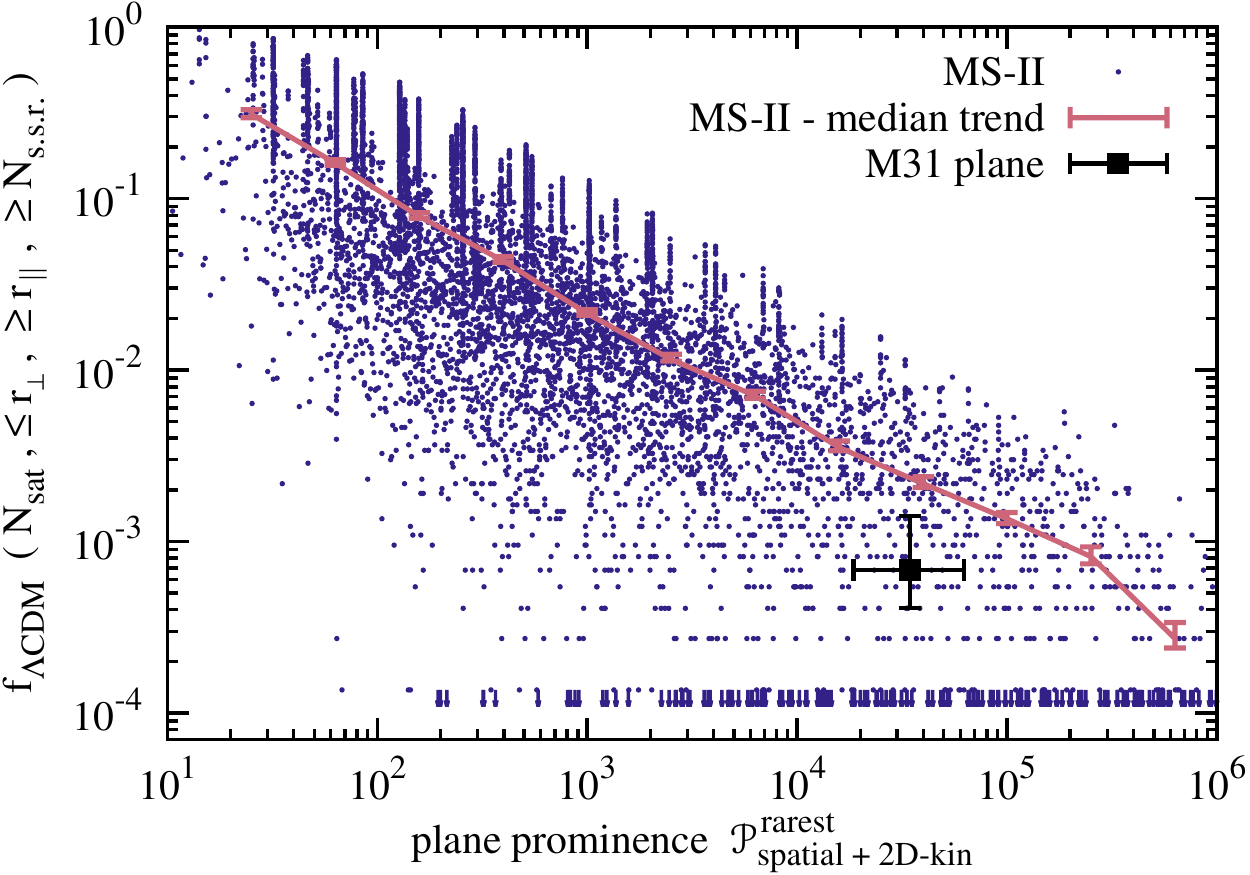}
     \caption{ The correlation between a plane's prominence, $\Pcombined{}$, and its incidence, $\fractLCDM$, among other \lcdm{} haloes. The small dots correspond to the rarest plane found for each \MII{} halo. The solid line shows the median trend. The large symbol with error bars shows the position of the M31 plane. The small down-pointing arrows found at the bottom of the graph show upper limits corresponding to planes that do not have another counterpart in \MII{}.
      }
     \label{fig:M31_fraction_haloes_with_similar_planes_scatter}
\end{figure} 


We define the incidence or frequency of a plane using the approach of \citet{Ibata2014b}. Each plane of satellites is characterized by: its number of members, $\Nsat$; how many of them share the same sense of rotation, $\Ncor$; the thickness, $\rPerp$; and radial extent, $\rAlong$, of the plane. Then, the frequency or incidence, $\fractLCDM$, of this plane is given by the fraction of \lcdm{} systems that have a similar plane. To describe the procedure, we exemplify it for the case of two systems A and B. We are interested in the frequency of the rarest plane of satellites of halo A and we wish to find out if halo B has a similar plane. We take all possible satellite configurations of system B that have $\Nsat^{\rm A}$ members, which we find using the procedure described in \refappendix{appendix:select_subset}. If any of those configurations is similar to plane of A, i.e. $\rPerp^{\rm B}\le\rPerp^{\rm A}$ and $\rAlong^{\rm B}\ge\rAlong^{\rm A}$ and $\Ncor^{\rm B}\ge\Ncor^{\rm A}$, then halo B has at least one planar configuration similar to that of system A. We compute the frequency for each \MII{} and \COCO{} halo, by taking the characteristics of the rarest plane identified around each halo. Each such rarest plane is compared to the satellites distribution of all the other \pandas{}-like mocks. This equates to comparing the rarest plane found in one halo with all possible planes, not only the rarest ones, around all other systems.

The frequency of the rarest \pTwo{} planes of satellites is shown in \reffig{fig:M31_fraction_haloes_with_similar_planes}, where we plot the CDF of the frequency for \MII{} and \COCO{} planes. We find that over half of \lcdm{} haloes have a plane with a frequency of $0.02$ or lower, and one tenth have a frequency as low as $0.001$. It is important to note that the low frequency is not a consequence of our plane identification method. For each halo, our method selects the plane that is the least likely to be a statistical fluctuations, which is fully independent of the planes found around other \lcdm{} haloes. Thus, there is a large diversity of planes of satellites. In other words, if we find a planar configuration around one system, it does not tell us anything about the properties of planes around other haloes.

\begin{table*}
    \begin{minipage}{\textwidth}
    
    \begin{center}
    \renewcommand{\arraystretch}{1.05}
    \caption{ The positions and velocities of the 11 classical MW satellites with respect to the Galactic Center.}
    \label{tab:MW_satellite_data}
    \begin{tabular}{ L{2cm} llL{2.2cm} lll }
        \hline
        Name &   $x (\rm kpc)$ &   $y (\rm kpc)$ &   $z (\rm kpc)$ &   $V_{x} (\rm km/s)$ &   $V_{y} (\rm km/s)$ &   $V_{z} (\rm km/s)$  \\[0.cm]
        \hline\hline
        Sagittarius    &  $17.1\pm1.9$ &  $2.5\pm0.2$    &  $-6.5\pm0.5$           &  $234\pm7$   &  $19\pm21$    &  $224\pm21$    \\
        LMC             &  $-0.5\pm0.3$ &  $-41.8\pm1.6$ &  $-27.5\pm1.1$         &  $-42\pm12$ &  $-226\pm13$ &  $234\pm16$    \\
        SMC             &  $16.5\pm1.6$ &  $-38.5\pm2.4$ &  $-44.7\pm2.8$         &  $2\pm18$     &  $-161\pm26$ &  $149\pm21$    \\
        Draco           &  $-4.4\pm0.3$ &  $62.3\pm4.9$   &  $43.2\pm3.4$           &  $74\pm24$   &  $43\pm14$    &  $-210\pm19$   \\
        Ursa Minor  &  $-22.2\pm0.6$ &  $52.0\pm2.1$  &  $53.5\pm2.1$           &  $7\pm28$     &  $89\pm20$    &  $-186\pm20$   \\
        Sculptor       &  $-5.2\pm0.3$  &  $-9.7\pm0.7$   &  $-85.3\pm5.9$         &  $-33\pm44$ &  $188\pm45$  &  $-99\pm6$       \\
        Sextans        &  $-36.6\pm1.3$  &  $-56.8\pm2.6$ &  $57.8\pm2.7$         &  $-168\pm160$ &  $114\pm133$  &  $117\pm127$   \\
        Carina          &  $-25.0\pm1.0$  &  $-95.8\pm5.5$ &  $-39.7\pm2.3$       &  $-74\pm44$ &  $8\pm19$  &  $40\pm41$   \\
        Fornax         &  $-41.1\pm2.7$  &  $-50.8\pm4.1$ &  $-134\pm11$         &  $-38\pm27$ &  $-156\pm42$  &  $113\pm18$   \\
        Leo II           &  $-77.3\pm4.1$  &  $-58.3\pm3.5$ &  $214\pm13$           &  $102\pm127$ &  $237\pm156$  &  $117\pm50$   \\
        Leo I            &  $-124\pm7$     &  $-119\pm7$      &  $192\pm12$           &  $-167\pm31$ &  $-35\pm33$  &  $96\pm24$   \\
        \hline
    \end{tabular}
    \end{center}
    
    \small  \smallskip
    The x-axis points from the Sun towards the Galactic Centre, the y-axis points in the direction of Galactic rotation at Sun's position and the z-axis points towards the North Galactic Pole. \MCn{Since this is a rotated coordinate system, the uncertainties are correlated and are very anisotropic in the plane of the sky.} 
    \end{minipage}
\end{table*}

We performed the same calculation for the M31 system. For each Monte Carlo realization, we compute the incidence of the rarest \pTwo{} plane of that realization by comparing with the \pandas{}-like mocks. We found that the M31 plane has a frequency of $6.8^{+7.2}_{-2.7}\times10^{-4}$ ($1\sigma$ confidence interval; see \reftab{tab:plane_incidence}), which is in good agreement with the results of previous studies \citep{Ibata2014b,Pawlowski2014c}. This low incidence of the M31 plane has been claimed by \citet{Ibata2014b} and \citet{Pawlowski2014c} to be a source of discrepancy with \lcdm{}. From \reffig{fig:M31_fraction_haloes_with_similar_planes}, which shows the M31 frequency as a vertical dashed line, we find that, within \lcdm{}, $5.1^{+4.5}_{-0.9}\%$ of systems have planes with even lower frequencies. Thus, the low incidence of the M31 plane does not pose a challenge to the current paradigm, in fact, it is consistent with \lcdm{} predictions.

\reffig{fig:M31_fraction_haloes_with_similar_planes_scatter} investigates which planes are the ones with the lowest incidence. For this, we plot the incidence of each plane as a function of its prominence, and find a strong anti-correlation between the two, albeit with a large scatter. The vertical concentrations of points correspond to planes that have a very high \pFour{} prominence but only a very low \pOne{} one, with the discrete nature of the \pFour{} test leading to many planes having very similar $\Pcombined{}$ values. So, on average, the more prominent a plane is, the lower is its incidence among \lcdm{} systems. This explains why the M31 plane, which has a high prominence, also has a low incidence. \MCn{The M31 plane, shown in \reffig{fig:M31_fraction_haloes_with_similar_planes_scatter} as a square symbol with error bars, is consistent with the object-to-object scatter in the prominence-incidence relation. Compared to the median trend, the M31 plane has an ${\sim}5$ times lower incidence for its prominence, which could be due to its unusually large radial extent (see bottom panel of \reffig{fig:M31_2d_histogram_Nsat}).}

\section{MW-like planes of satellites}
\label{sec:MW_planes}
In this section we investigate the MW system of satellites and how it compares with other \lcdm{} planar configurations. Compared to the M31 analysis, there are three main differences: we use systems with 11 instead of 27 satellites, the survey geometry is different and, most importantly, we can perform additional tests since we have proper motion data for the MW satellites. In this analysis, we consider only the brightest 11 classical Galactic satellites since only these objects have measured proper motions. A twelfth satellite, Canes Venitici, has a similar absolute magnitude as the faintest of the classical satellites \citep{McConnachie2012}, but it does not have a measured proper motion, so we do not include it in this study.

To create MW-like mocks, we consider the 11 satellites with the largest stellar masses that are within a distance of $260\kpc$ from the central galaxy (corresponding to Leo I which is the furthest at $254\kpc$) and that, at the same time, are outside an obscuration region consisting of $33\%$ of the sky. The obscuration region accounts for the Galactic zone of avoidance, where, due to large extinction and confusion by foreground stars, it is possible to have yet undiscovered bright satellites. For this, we use the estimate of \citet{Willman2004} according to which the census of Galactic dwarf galaxies may be $33\%$ incomplete. \citet{Yniguez2014} estimates an even higher incompleteness, with most of those undetected systems further than $100\kpc$ from the Galactic Centre. A satellite is lying inside the obscuration region, and hence undetected, if its latitude, $\theta$, is in the range $-\theta_{\rm crit}\le \theta\le \theta_{\rm crit}$, with $\theta_{\rm crit}=19.5^\circ$ (due to an error, \citet{Wang2013} misquoted the critical angle as having a value of $9.5^\circ$, which would correspond to an obscuration region of $17\%$). To generate mock observations, we take three viewing angles for each simulated halo such that the mock north Galactic pole corresponds to the x, y and z axes of the simulation. This procedure yields 8547 mock satellite systems for \MII{} and 189 for \COCO{}.

\subsection{The MW system}
In a first step, we analyse the planar configurations found around the MW. For each of the 11 Galactic satellites, we take their radial distance and velocity, as well as the errors associated with these quantities, from \citet{McConnachie2012}. To compute the mean proper motion, we follow the approach of \citet{Pawlowski2013b} and weigh the different measurements according to their errors. We obtain the same mean proper motions as they do, except for Draco, where the latest measurement, $(\mu_\alpha,\mu_\delta)=(0.177\pm0.063, -0.221\pm0.063)\marcsyear$ \citep{Pryor2015}, is significantly different from the value given by \citet{Pawlowski2013b}. Using this updated measurement results in a weighted mean value $(\overline{\mu}_\alpha,\overline{\mu}_\delta)=(0.187\pm0.063,-0.201\pm0.063)\marcsyear$. We also included an additional proper motion measurement for Sagittarius of $(\mu_\alpha,\mu_\delta)=(-2.95\pm0.21, -1.19\pm0.16)\marcsyear$ \citep{Massari2013} that resulted in a mean value of $(\overline{\mu}_\alpha,\overline{\mu}_\delta)=(-2.711\pm0.066,-1.043\pm0.065)\marcsyear$, nearly the same as the mean value used by \citet{Pawlowski2013b}.

We transform the satellite positions and velocities to a Cartesian coordinate system with the origin at the Galactic Centre. The x-axis points from the Sun towards the Galactic Centre, the y-axis points in the direction of Galactic rotation at the Sun's position and the z-axis points towards the North Galactic Pole. For this transformation we adopt: the distance of the Sun from the Galactic Centre $d_\sun = 8.29\pm0.16\kpc$, the circular velocity at the Sun's position, $V_{\rm circ} = 239\pm5\kms$ \citep{McMillan2011}, and the Sun's motion with respect to the Local Standard of Rest, $(U,V,W)=(11.1\pm0.8,12.2\pm0.5,7.3\pm0.4)\kms$ \citep{Schonrich2010}. To account for observational errors, we generate $1000$ Monte Carlo realizations of the MW system of satellites. We sample the satellite positions and proper motions from Gaussian distributions centred on the most likely values of each quantity and with dispersion equal to the uncertainties. Similarly, we also account for the errors in $d_\sun$, $V_{\rm circ}$ and the Local Standard of Rest. Following this, we transform from heliocentric coordinates to Galactic ones, with the result used as input for our plane detection method. We summarize in \reftab{tab:MW_satellite_data} the mean positions, velocities and $1\sigma$ errors associated with each Galactic satellite.

As for the M31 case, for each Monte Carlo realization of the MW system we compute the rarest \pOne{}, \pTwo{} and \pThree{} planes. The results are summarized in \reftab{tab:plane_properties}. Independently of the plane type, we find that the rarest plane is the one that contains all the 11 Galactic satellites. This is in agreement with previous studies that found that all the classical satellites are members of the MW satellite plane \citep[e.g.][]{Kroupa2005}.

\subsection{The rarest MW-like planes}
\begin{figure}
     \centering
     \includegraphics[width=1.03\linewidth,angle=0]{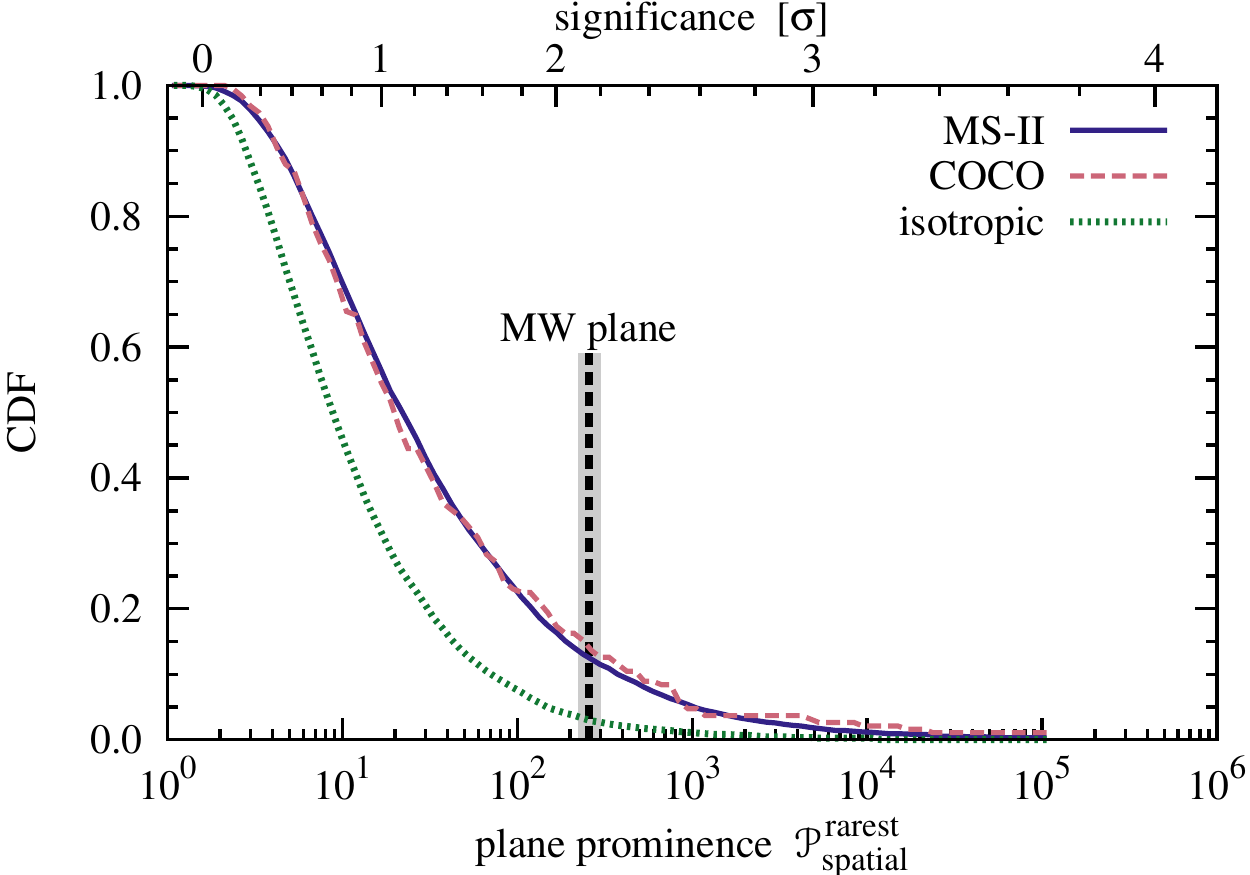}
     \caption{ The CDF of the prominence, $\Pspatial{}$, of the rarest \pOne{} plane of satellites for mock MW observations. The vertical dashed line and shaded region show the prominence and $1\sigma$ error for the MW plane of satellites, with $(12\pm1)\%$ of \lcdm{} haloes having a more prominent plane. }
     \label{fig:MW_spatial_plane_PDF}
     
     \includegraphics[width=1.03\linewidth,angle=0]{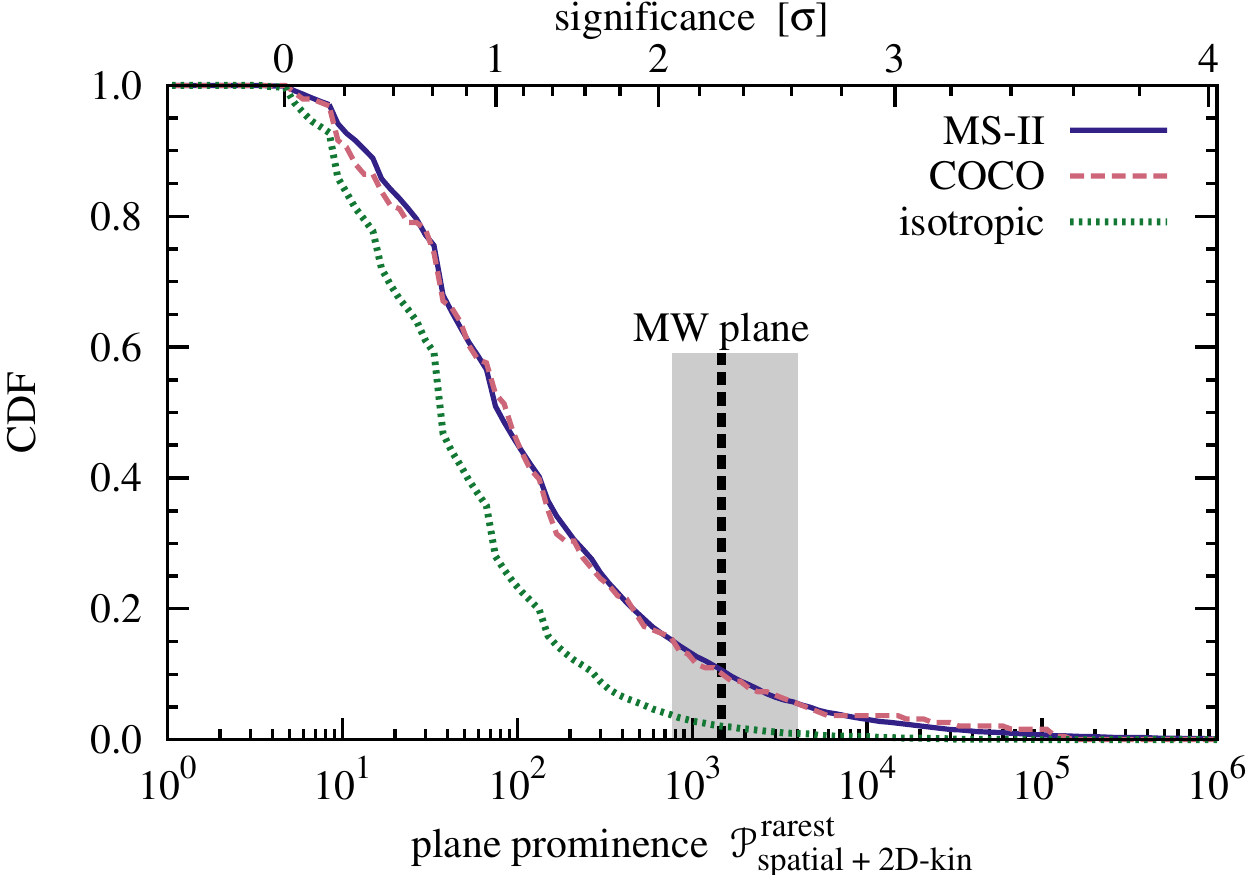}
     \caption{ As \reffig{fig:MW_spatial_plane_PDF}, but for the prominence, $\Pcombined{}$, of the rarest \pTwo{} plane of satellites. In this case, $(10\pm5)\%$ of \lcdm{} hosts have a more prominent plane than the MW. }
     \label{fig:MW_combined_plane_PDF}
     
     \includegraphics[width=1.03\linewidth,angle=0]{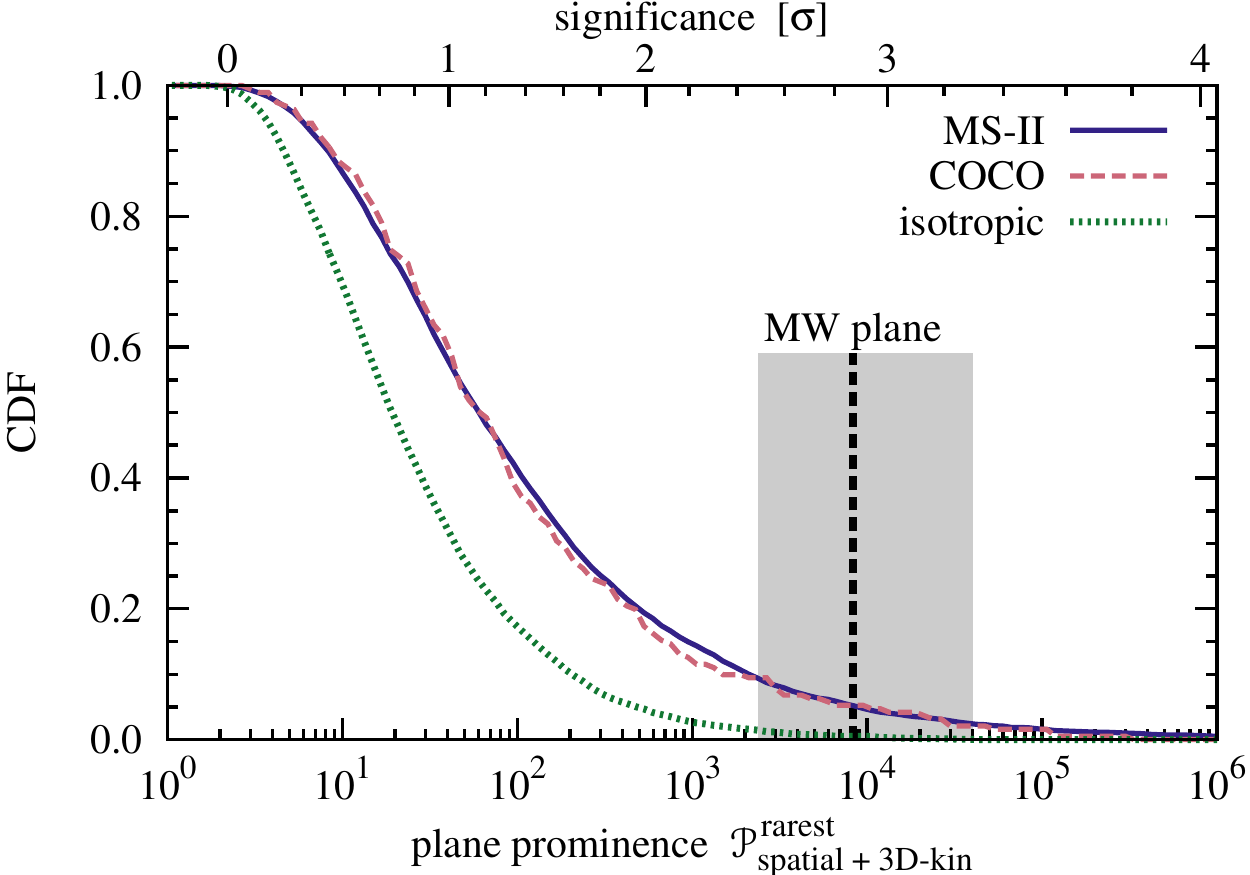}
     \caption{ As \reffig{fig:MW_spatial_plane_PDF}, but for the prominence, $\PcombinedMW{}$, of the rarest \pThree{} plane of satellites. For this test, $5.0^{+4.1}_{-2.7}\%$ of \lcdm{} haloes have a more prominent plane than the MW. }
     \label{fig:MW_combined_orbit_plane_PDF}
\end{figure}

To put the MW plane of satellites into context, we proceed by identifying the rarest planes around mock MW-like observations. The outcome is shown in \reffigs{fig:MW_spatial_plane_PDF}{fig:MW_combined_orbit_plane_PDF} that give the CDF of the prominence of the rarest \pOne{}, \pTwo{} and \pThree{} planes, respectively. The main conclusions are the same as for the \pandas{} mocks analysed in \refsec{subsec:M31_rarest_planes}.

The look-elsewhere effect again plays an important role for the MW-like mocks, with \pOne{} planes having a trial factor ${\sim}8$, while \pTwo{} and \pThree{} planes have a trial factor ${\sim}30$. The trial factors are roughly constant in the region where the isotropic CDF is below unity, reminiscent of the results in \reffig{fig:M31_trial_factors}. For brevity, we do not show these results. The trial factors are lower in the case of the MW mocks than in the case of M31, reflecting the narrower range used to search for planar configurations in the MW (11 satellites compared to 27 for the M31).

\reffigs{fig:MW_spatial_plane_PDF}{fig:MW_combined_orbit_plane_PDF} also indicate the prominence of the MW plane of satellites as a dotted vertical line. The MW plane stands out the most in terms of its \pThree{} prominence since in this case it corresponds to a $2.8\sigma$ statistical fluctuation. Not accounting for the look-elsewhere effect, would lead one to estimate the MW plane as a $3.8\sigma$ detection. While the MW \pThree{} plane is quite conspicuous, it is consistent with \lcdm{} since $5.0^{+4.1}_{-2.7}\%$ of galactic-mass haloes have even more prominent planes. 


\MCn{As we emphasized in \refsec{subsec:defining_spatial_and_orbital_planes}, one needs to be careful when interpreting the results of the \pFive{} analysis since this test has been designed \textit{a posteriori}. In fact, one could easily come up with other 3D kinematic tests that are physically better motivated, as we discussed in \refsec{subsec:defining_spatial_and_orbital_planes}. Given that observationally we have only one satellite system with 3D velocities, it is impossible at present to assess if the \pFive{} test that we have applied is generic and thus appropriate to the whole population of satellite systems or if it is matched to the particular details of the MW satellites. If the latter is true, then the fraction of \lcdm{} haloes with more prominent planes does not convey any physically meaningful information. Thus, there is currently not enough data to decide if one should be concerned that only ${\sim}5\%$ of \lcdm{} haloes have a more prominent \pThree{} plane than the MW one.}

\begin{figure}
     \centering
     \includegraphics[width=1.0\linewidth,angle=0]{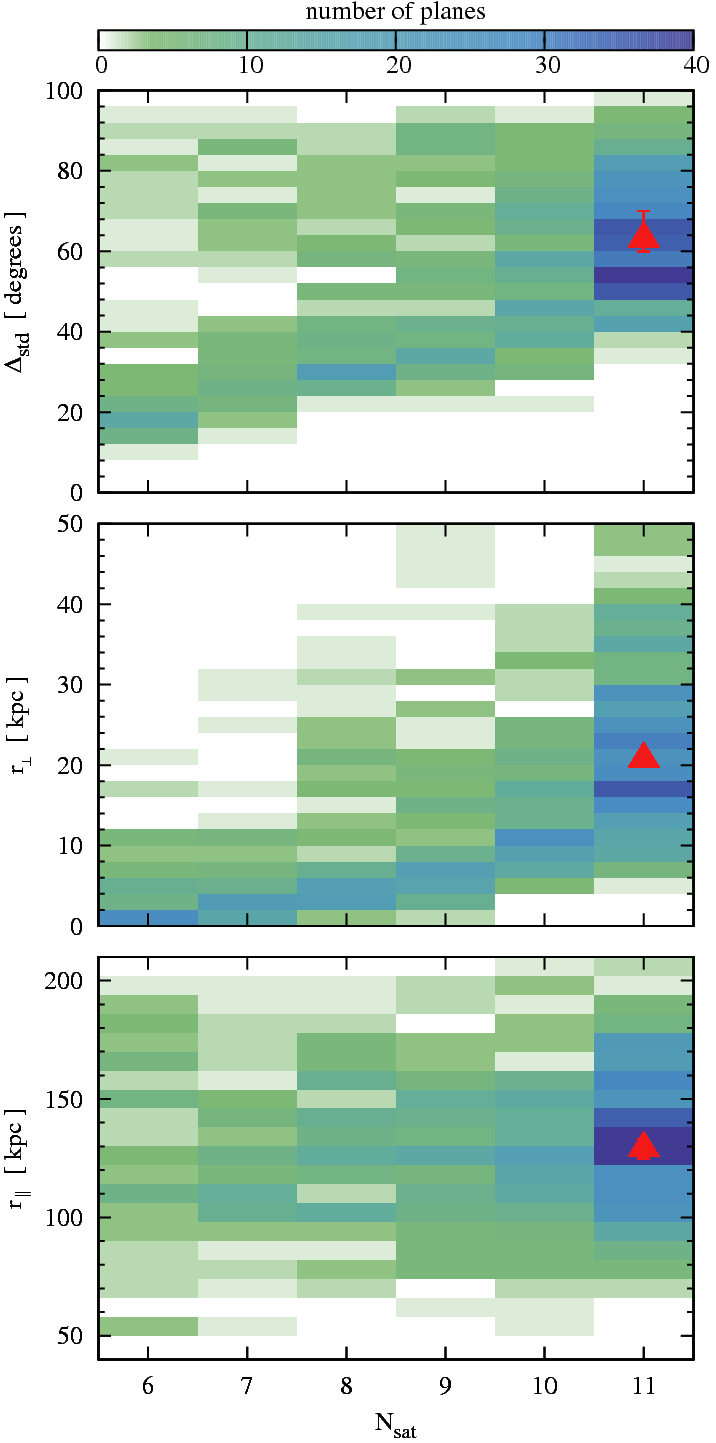}
     \caption{ The characteristics of the rarest planes of satellites that are at least as prominent as the MW plane of satellites, i.e. $\PcombinedMW{} \ge 2.3\times 10^{4}$. There are 778 such systems. The grid cells are coloured according to the number of systems with those properties. The three panels show the orbital pole dispersion, $\Dorbit$ (top), the thickness, $\rPerp$ (centre), and the radial extent of the plane, $\rAlong$ (bottom), as a function of the number of satellites in the plane, $\Nsat$. The large triangle shows the corresponding characteristics of the MW plane of satellites. }
     \label{fig:MW_2d_histogram_Nsat}
\end{figure} 

In \reffig{fig:MW_2d_histogram_Nsat} we plot the properties of the rarest \pThree{} planes that are at least as prominent as the MW satellite plane. In analogy to \refsec{subsec:M31_plane_incidence}, we find that the planes are characterized by a large diversity in orbital pole dispersion, thickness and radial extent. A plane can be very prominent by being very thin, by having a small orbital pole dispersion or by a combination of the two, which explains the large scatter seen in the $\Dorbit$ and $\rPerp$ properties. Interestingly, we find that most of such planes ($43\%$) have $\Nsat=11$, the same as the number of members in the MW plane of satellites whose characteristics are shown as a large triangle.

\subsection{The incidence of MW-like planes}
\label{subsec:MW_plane_incidence}
\begin{figure}
     \centering
     \includegraphics[width=1.03\linewidth,angle=0]{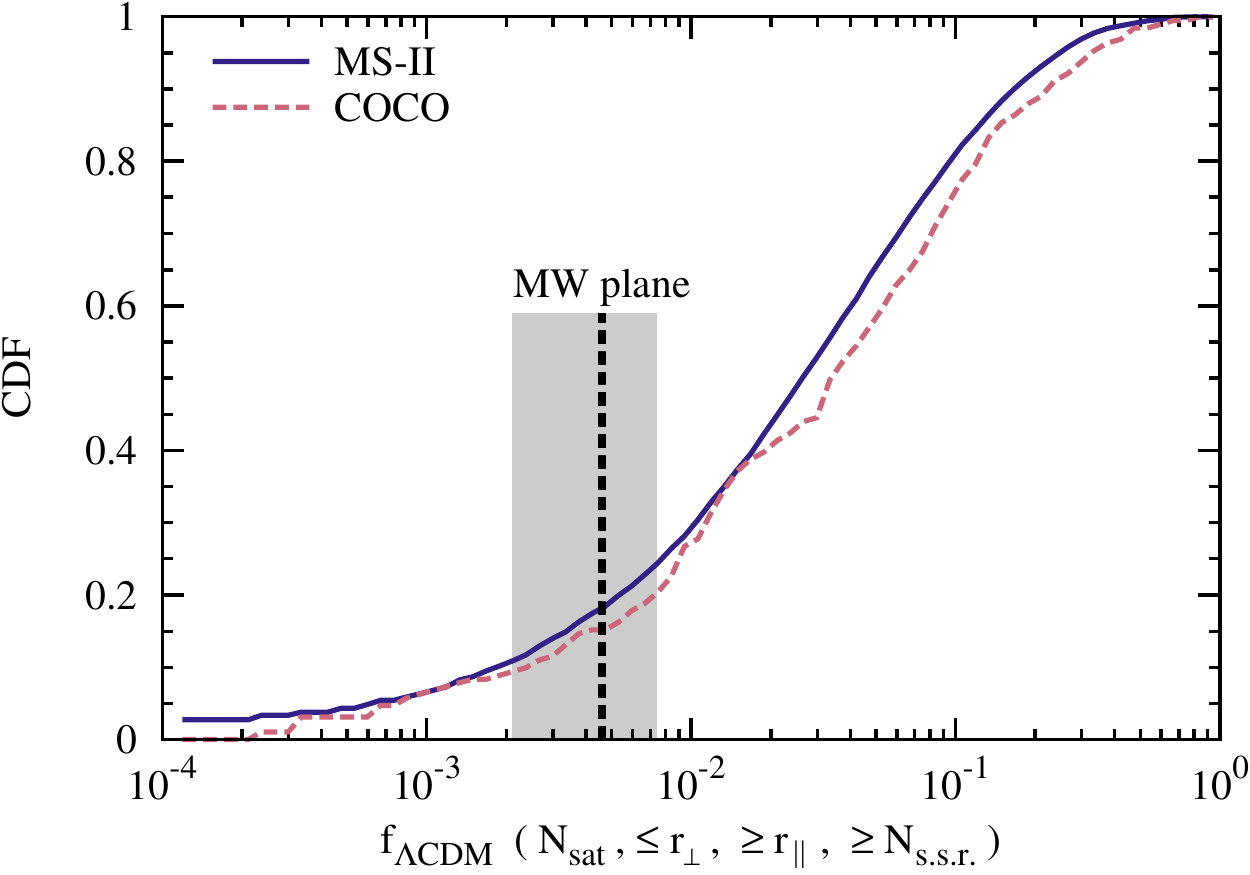}
     \caption{ The CDF of the incidence, $\fractLCDM$, of finding similar planes in \lcdm{}. As in the M31 case, this illustrates the diversity of planar configurations found in MW-like mocks. The \MII{} and \COCO{} results are consistent with the scatter expected due the low number of \COCO{} haloes. The vertical dashed line and the grey area show the incidence and the $1\sigma$ error for the MW plane of satellites. We find that $18^{+6}_{-8}\%$ of \lcdm{} systems have an even lower frequency than the MW plane. }
     \label{fig:MW_fraction_haloes_with_similar_planes_1}
     
     \includegraphics[width=1.03\linewidth,angle=0]{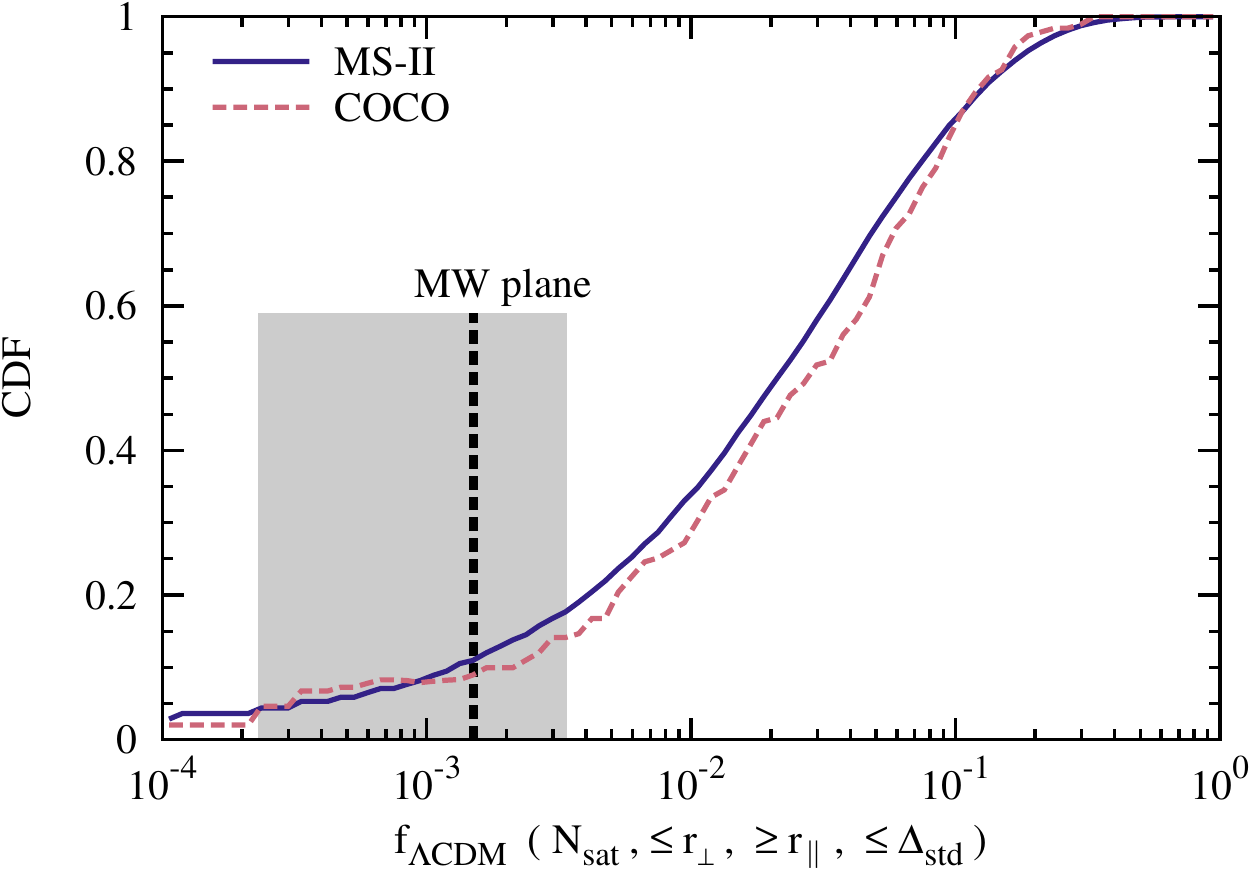}
     \caption{ As \reffig{fig:MW_fraction_haloes_with_similar_planes_1}, but for the incidence, $\fractLCDMorbit$. In this case, we compare the orbital pole dispersion, $\Dorbit{}$, between different planes and not the number of satellites sharing the same sense of rotation, $\Ncor{}$, as in \reffig{fig:MW_fraction_haloes_with_similar_planes_1}. We find that $11^{+6}_{-7}\%$ of \lcdm{} systems have an even lower incidence than the MW plane. Thus, the low incidence of the MW plane is not a symptom of discrepancy with \lcdm{}, as claimed by \citet{Pawlowski2014c}. }
     \label{fig:MW_fraction_haloes_with_similar_planes_2}
\end{figure} 

We follow the same approach as in \refsec{subsec:M31_plane_incidence} and compute the incidence of the rarest plane around each \lcdm{} halo. The outcome is presented in \reffig{fig:MW_fraction_haloes_with_similar_planes_1} which shows that most systems have planar configurations that are very infrequent, with over half of the haloes having a plane with an incidence of $0.03$ or lower. The slight disagreement between the \MII{} and the \COCO{} results is consistent with the scatter expected for the much smaller sample of \COCO{} systems. This appears as a systematic shift due to the correlations between points in the CDF. For this test, the MW plane of satellites has a frequency of $4.6^{+2.8}_{-2.5}\times10^{-3}$ (vertical dashed line in \reffig{fig:MW_fraction_haloes_with_similar_planes_1}) that is consistent with \lcdm{} expectations, since $18^{+6}_{-8}\%$ of haloes have an even lower incidence. 

Since for the Galactic satellites we have full kinematical data, we can define a new incidence, $\fractLCDMorbit$, that includes the 3D kinematics. This is similar to the incidence introduced in \refsec{subsec:M31_plane_incidence}, except that now we compare $\Dorbit{}$ between different planes instead of $\Ncor{}$.
The CDF of this new frequency is shown in \reffig{fig:MW_fraction_haloes_with_similar_planes_2}. Compared to \reffig{fig:MW_fraction_haloes_with_similar_planes_1}, the outcome is very similar except for a slight shift in the CDF towards the left, i.e. towards lower frequencies. 

The new statistics indicates that the MW plane of satellites has an incidence of $1.5^{+1.9}_{-1.2}\times10^{-3}$, as shown by the vertical dashed line in \reffig{fig:MW_fraction_haloes_with_similar_planes_2}. This value agrees with the result of \citet[][entry 12 of their Table 5]{Pawlowski2014c}, whose test is very similar to ours, except that those authors considered the orbital pole dispersion of only 8 out of 11 satellites. Pawlowski et al claimed that this low incidence of the MW plane of satellites is indicative of a shortcoming of the \lcdm{} paradigm. Instead, we find that the low incidence indicates that planes of satellites are very diverse. The distribution of satellites around the MW agrees with \lcdm{} predictions since $11^{+6}_{-7}\%$ of similar mass haloes have an even lower frequency.

\section{Discussion and conclusions}
\label{sec:conclusion}
In this paper, we have investigated the prevalence and properties of planar configurations of satellites around galactic mass haloes. Using two very high resolution cosmological simulations, \MII{} and \COCO{}, we have built mock galaxy catalogues corresponding to the satellite distributions around the MW and M31. MW-like mocks consist of the most massive 11 satellites found outside an assumed zone of avoidance, mimicking the observations of the 11 classical Galactic satellites. M31-like mocks are modelled according to the \pandas{} footprint and consist of the most massive 27 satellites found within that region, corresponding to the M31 satellites observed by \pandas{}. 

We identify the most prominent planar configuration as the subsample of satellites whose spatial and kinematical distribution is the least likely to be a statistical fluctuation. Applying our approach to the MW and M31 observations results in the same planar distributions as determined by \citet{Kroupa2005} and \ibata{}, respectively, even though those studies used different identification methods. The good agreement is possibly due to \emph{a posteriori} selection bias, since those authors may have inadvertently tuned their methods to maximize the significance of the detection. This would result in all the methods converging to the same planes.

We have found that planar configurations of satellites are very common around \lcdm{} haloes, and, moreover, approximately $5$ and $9\%$ of haloes have even more prominent planes than those found in the MW and M31, respectively (see \reftab{tab:plane_properties}, and \reffigS{fig:M31_combined_plane_PDF}{fig:MW_combined_orbit_plane_PDF}). The \emph{look-elsewhere effect} is crucial in assessing the detection significance of a planar distribution, i.e. in estimating the probability of obtaining such a structure in an isotropic distribution  (\refsec{subsec:M31_look_elsewhere}). By neglecting this effect, one can easily overestimate the significance level by factors of ${\sim}30$ and ${\sim}100$ for the MW and M31 planes respectively. For example, while the M31 plane was originally reported to have a $99.998\%$ significance ($4.3\sigma$ detection; \ibata{}), accounting for the \emph{look-elsewhere effect} results in a more modest $99.7\%$ significance ($2.9\sigma$ detection).

While ubiquitous, the planes of satellites show a large diversity in characteristics, e.g. in the number of members, the plane thickness and radial extent, as well as the kinematical structure (see \reffigS{fig:M31_2d_histogram_Nsat}{fig:MW_2d_histogram_Nsat}). Most planar configurations are distinct, which has two major implications. Firstly, the notion of a representative plane of satellites does not exist since one cannot find a majority of \lcdm{} haloes that have the same planar configuration. Secondly, the large diversity of planes precludes using one or two observed systems for testing the cosmological paradigm on small scales. For such a test, a large sample of satellite systems would be needed to obtain a statistical measure of the system-to-system variation.

The diversity of the planes of satellites is also the root cause behind previous claims that planes found in observations are inconsistent with \lcdm{} \citep{Ibata2014b,Pawlowski2014c}. These authors computed the incidence of the MW and M31 plane of satellites to find out that roughly only 1 out of 1000 \lcdm{} systems have such planes. To understand this result, we have computed the incidence of the rarest plane of satellites identified around each \lcdm{} halo. We have found that the majority of planar configurations have a very low incidence and that $11$ and $5\%$ of \lcdm{} planes have even lower incidence than that of the planes of satellite in the MW and M31, respectively (see \reftab{tab:plane_incidence}, and \reffigS{fig:M31_fraction_haloes_with_similar_planes}{fig:MW_fraction_haloes_with_similar_planes_2}). The low incidence is a manifestation of the diversity of satellite planes and, thus, contrary to the claim by \citet{Pawlowski2014c}, it does not rule out the \lcdm{} paradigm.

While the planes of satellites around the MW and M31 are consistent with \lcdm{}, both systems fall in the $10\%$ tail of the distribution. If both planes were independent of each other, one might argue that there is only a ${\sim}1\%$ chance that both systems are randomly drawn from a \lcdm{} distribution. This interpretation is problematic for at least two reasons. Firstly, both the MW and the M31 are located in the same large-scale environment, which in turn determines the preferential directions of satellite accretion \citep{Libeskind2014,Libeskind2015}. Thus, if the environment is especially conducive to the formation of prominent satellite planes, then it may not be surprising that both systems host prominent planes. Secondly, the tests used to assess the prominence of these planes were designed \emph{a posteriori}, after investigating the observational data. This is especially true for the definition of the orbital pole dispersion that has been motivated by examining the MW data. Such an \emph{a posteriori} approach incurs the danger of designing tests that are specifically matched to the peculiarities of a particular system and are not characteristic of the population as a whole.


Our analysis has shown that the planes of satellites identified in the MW and M31 are consistent with \lcdm{} predictions based on high resolution cosmological simulations. This agrees with the results of \citet{Cautun2015a}, which compared the spatial and kinematical distributions of satellites around a large sample of isolated galaxies in SDSS to find agreement between observations and theoretical predictions. Previous claims of an inconsistency with \lcdm{} are based on misinterpreting the low incidence of satellite planes \citep[e.g.][]{Ibata2014b,Pawlowski2014c} and on non-robust detections \citep[][see \refsec{sec:introduction} and \citealt{Cautun2015a} for details]{Ibata2014}.
Thus, there is no convincing evidence for a discrepancy between observed planes of satellites and the \lcdm{} predictions.

\section*{Acknowledgements} 
We are grateful to Julio Navarro and Simon White for helpful discussions and comments.
We also thank the anonymous referee for their comments that have helped us improve the paper.
This work was supported in part by ERC
Advanced Investigator grant COSMIWAY [grant number GA 267291] and the
Science and Technology Facilities Council (STFC) [grant number ST/F001166/1,
ST/I00162X/1].
SB is supported by STFC through grant [ST/K501979/1,ST/L00075X/1].
QG acknowledges support from the Strategic Priority Research Program ``The Emergence of Cosmological Structure” of the Chinese Academy of Sciences (No. XDB09000000) and the ``Recruitment Program of Global Youth Experts'' of China. 
WW is supported by JRF grant number RF040353.  
This work used the DiRAC Data Centric system at Durham University, 
operated by ICC on behalf of the STFC DiRAC HPC Facility
(www.dirac.ac.uk). This equipment was funded by BIS National
E-infrastructure capital grant ST/K00042X/1, STFC capital grant
ST/H008519/1, and STFC DiRAC Operations grant ST/K003267/1 and Durham
University. DiRAC is part of the National E-Infrastructure.  This
research was carried out with the support of the ``HPC Infrastructure
for Grand Challenges of Science and Engineering'' Project, co-financed
by the European Regional Development Fund under the Innovative Economy
Operational Programme.
The \COCO{} simulation has been run at the supercomputer centre of
the Interdisciplinary Centre for Mathematical and Computational Modelling at University of Warsaw.

\hypertarget{labelHypertarget}{}

\newcommand{\jcap}{JCAP} 
\newcommand{\pasa}{PASA}
\bibliographystyle{mn2e}
\bibliography{andromeda_reference}

\appendix

\section{Plane identification}
\label{appendix:overall}
Here we present the practical implementation of the plane identification procedure.

\subsection{Selecting subsets of satellites}
\label{appendix:select_subset}
We first describe how we identify the interesting subsets of satellites, which, in the next step, are used to find the rarest planar configurations. The simplest approach would be to take into account every possible combination of $\Nsat$ satellites out of a maximum of $\Nmax$ objects, with $3\le\Nsat\le\Nmax$. Planes with 2 or fewer objects are not physically interesting since any two satellites will determine a plane of thickness $\rPerp=0$. This naive approach, however, would result in a very large number of combinations that need to be considered, since for fixed $\Nsat$ the number of unique combinations is
\begin{equation}
    \frac{\Nmax !}{\Nsat!(\Nmax-\Nsat) !} \;.
\end{equation}
In the case of the M31 system, we have $\Nmax=27$ satellites, so choosing $\Nsat=14$ would result in $2\times10^{7}$ subsets that need to be considered. This analysis would have to be done for many thousands of systems, and for each we would need to generate $10^{5}$ isotropic distributions. Such an approach is not feasible in practice.

To overcome the immense computational challenge described above, we consider only configurations in which the plane members are the closest satellites to the plane. Thus, no other galaxy can be found closer to the plane than the furthest plane member. This is in line with the plane definitions used in earlier studies \citep[e.g.][]{Bahl2014, Gillet2015}. 
We start by selecting a sample of $N$ planes centred on the host galaxy and characterized by the normal vector, $\mathbf{n_{\rm plane}}$. To obtain these planes, we generate normal vectors that are uniformly distributed on half a sphere, since the opposite hemisphere corresponds to identical planes flipped upside down. For each such plane, we order the $\Nmax$ satellites according to their distance to the plane. The interesting subsets of satellites are those made of the closest $3,\; 4,\;\ldots,\;\Nmax$ objects from each plane.

To make sure that we identify all possible satellite subsets, we would like to have a very large number of random planes, $N$. In turn, increasing $N$ incurs a significantly larger computational cost and ends up adding mostly duplicate subsets of satellites, which were already identified for small values of $N$. We found the best compromise to be $N=10^3$, which is large enough to contain a significant fraction of all possible subsets. Using $N=10^3$ we find $93\%$ ($70\%$) of the subsets we would identify using $N=10^4$ for $\Nmax=11$ ($27$), which corresponds to the total number of satellites in the MW (M31) system. This means that for some systems we are missing the satellite subset corresponding to the rarest plane. In those cases we end up identifying the second or the third rarest planes as the most prominent planar configurations. 
Using a small sample of around $200$ \lcdm{} haloes, we have checked that using $N=10^4$ instead of $N=10^3$ brings only minor changes to the CDF of the prominence, $\Prarest$, of the rarest plane (e.g. \reffig{fig:M31_spatial_plane_PDF}) and to the CDF of the frequency, $\fLCDM{}$, of those planes (e.g. \reffig{fig:M31_fraction_haloes_with_similar_planes}). Thus, any missing subsets of satellites will not change our overall conclusion.

The subsets of satellites used to compute the frequency of the rarest planes (\refsecs{subsec:M31_plane_incidence}{subsec:MW_plane_incidence}) were identified employing the same procedure except that we used $N=10^5$. That is, we used $10^5$ random planes uniformly distributed on a hemisphere, which is the same as the approach used by \citet{Bahl2014} and subsequent studies.

\subsection{Generating isotropic distributions}
\label{appendix:isotropic_distribution}
Each isotropic realization is generated by picking random polar and azimuthal angles\footnote{The cosine of the polar angle and the azimuthal angle are selected from a uniform distribution spanning the interval $[-1,1]$ and $[0,2\pi]$, respectively.}
for each satellite, while keeping constant the radial distance from the host galaxy\footnote{Each random realization lies within the survey mask, which is \pandas{} for the M31 and a $19.5^\circ$ obscuration angle for the MW. If a random point falls outside the mask, we generate new random angles till the point is located within the survey mask.}. Thus, each isotropic realization has the same radial distribution of satellites as the original system. This point is crucial, since the radial distribution of satellites has a strong effect on the thickness of the resulting planes. Radially concentrated satellite distributions result in thinner planes than more radially extended ones. Thus, we need to generate new isotropic realizations for each system of satellites.

When constructing the isotropic distributions, we also generate random 3D velocities, which are used for computing the distribution of orbital pole dispersions, $\Dorbit$. Since we are only interested in the direction of the orbital momentum, the magnitude of the velocity is not important. Thus, the velocities are generated by picking random polar and azimuthal angles for each satellite, with the two angles fully independent from the random polar and azimuthal angles used to obtain the position of each satellite.

\subsection{The probability distribution of statistical fluctuations}
\label{appendix:probability}
We now describe how to estimate the probability that the spatial or kinematical distribution of a set of $\Nsat$ satellites is the result of a statistical fluctuation. This probability is computed using isotropic distributions, which characterize the degree of planarity expected from chance alignments and from the discreteness of the satellite distribution.

\begin{figure}
     \centering
     \includegraphics[width=1.01\linewidth,angle=0]{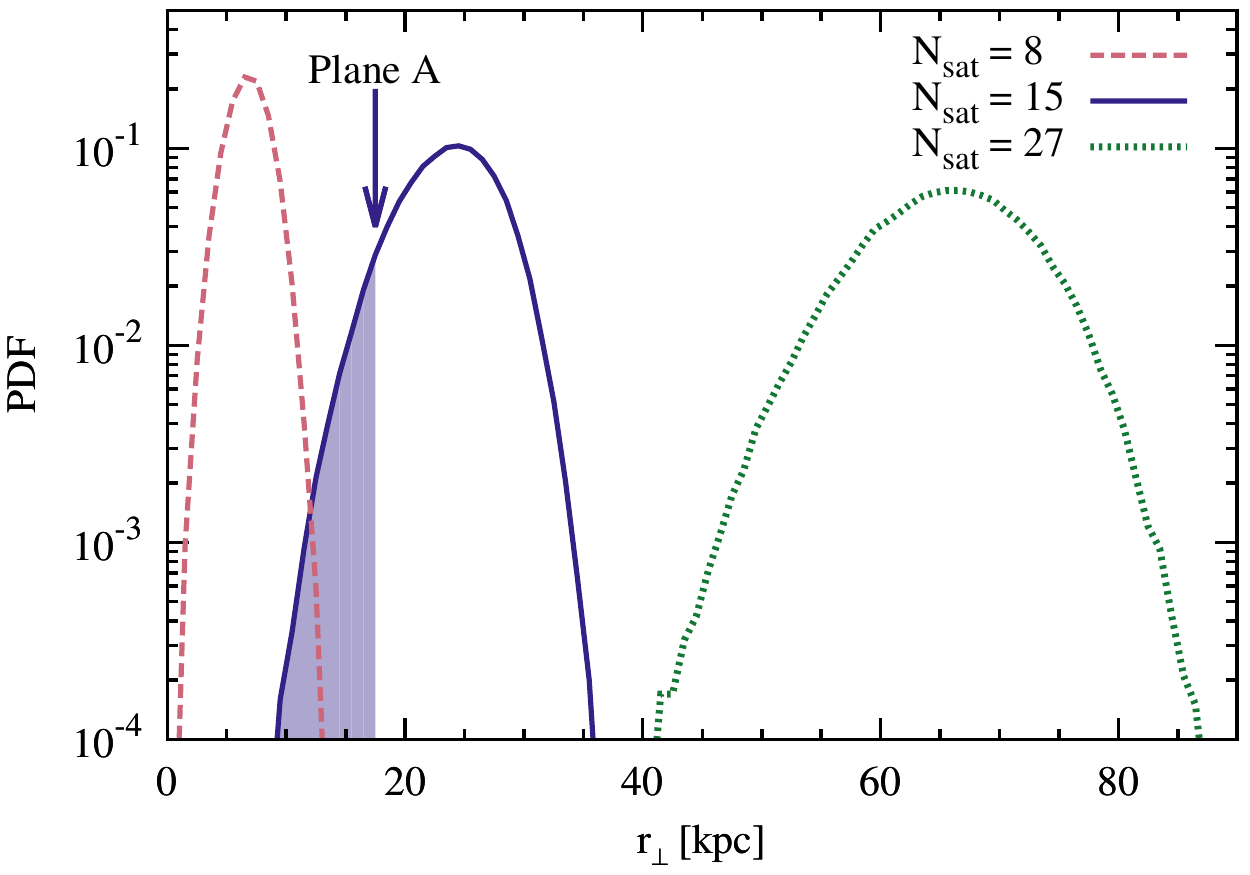}
     \caption{ The PDF of the plane thickness, $\rPerp$, for isotropic satellite distributions inside the \pandas{} survey footprint. We show results for planar configurations that contain $\Nsat=8$, 15 and 27 satellites out of a maximum of $\Nmax=27$ satellites. The vertical arrow and shaded region illustrate the probability that a fictitious plane A, which has $\Nsat=15$ and $\rPerp{}=17.5\kpc$, is due to a statistical fluctuation (see the text for details). }
     \label{fig:definition_plane_rms_width}
\end{figure}

The probability of obtaining by chance a plane of $\Nsat$ that is thinner than  $\rPerp{}$ is given by
\begin{equation}
     p\left( \le \rPerp \;| \; \Nsat \right) = \int_{0}^{\rPerp} PDF^{\rm isotropic}_{\rm spatial;\;\Nsat}(r'_{\perp}) \;\mathrm{d} r'_{\perp}
    \label{eq:p_spatial_one_plane_2} \;,
\end{equation}
where the integrand is the PDF of obtaining in an isotropic distribution planes with $\Nsat{}$ members and thickness, $r'_{\perp}$. To compute the PDF, for each halo we generate $10^5$ isotropic realizations using the procedure described in \refappendix{appendix:isotropic_distribution}. For each such realization we find the thinnest plane with $\Nsat{}$ members. The corresponding histogram over all realizations gives the PDF of $r'_{\perp}$ values. The resulting PDF, for the case of the M31 system ($\Nmax=27$ satellites), is shown in \reffig{fig:definition_plane_rms_width}. For clarity, we only give the planes with $\Nsat=8$, 15 and 27 members. The figure also illustrates, in an intuitive fashion, the meaning of \eq{eq:p_spatial_one_plane_2}. We exemplify this using a fictitious plane A that contains $\Nsat=15$ members and whose thickness is shown with a vertical solid arrow.  \eq{eq:p_spatial_one_plane_2} corresponds to the shaded area to the left of the solid arrow.

The probability of obtaining by chance a configuration of $\Nsat$ satellites in which at least $\Ncor{}$ members share the same sense of rotation is given by the binomial distribution with a success probability of 0.5 . Thus,
\begin{equation}
     p\left( \ge \Ncor \;|\; \Nsat \right) = 2 \; \frac{\Nsat!}{\Ncor!(\Nsat-\Ncor)!} \; 2^{\Nsat}
    \label{eq:isotropic_PDF_corotation} \;,
\end{equation}
where the first factor of $2$ comes from the fact that we do not fix a preferential sense of rotation, allowing both clockwise and counter-clockwise rotations.
 
\begin{figure}
     \centering
     \includegraphics[width=1.0\linewidth,angle=0]{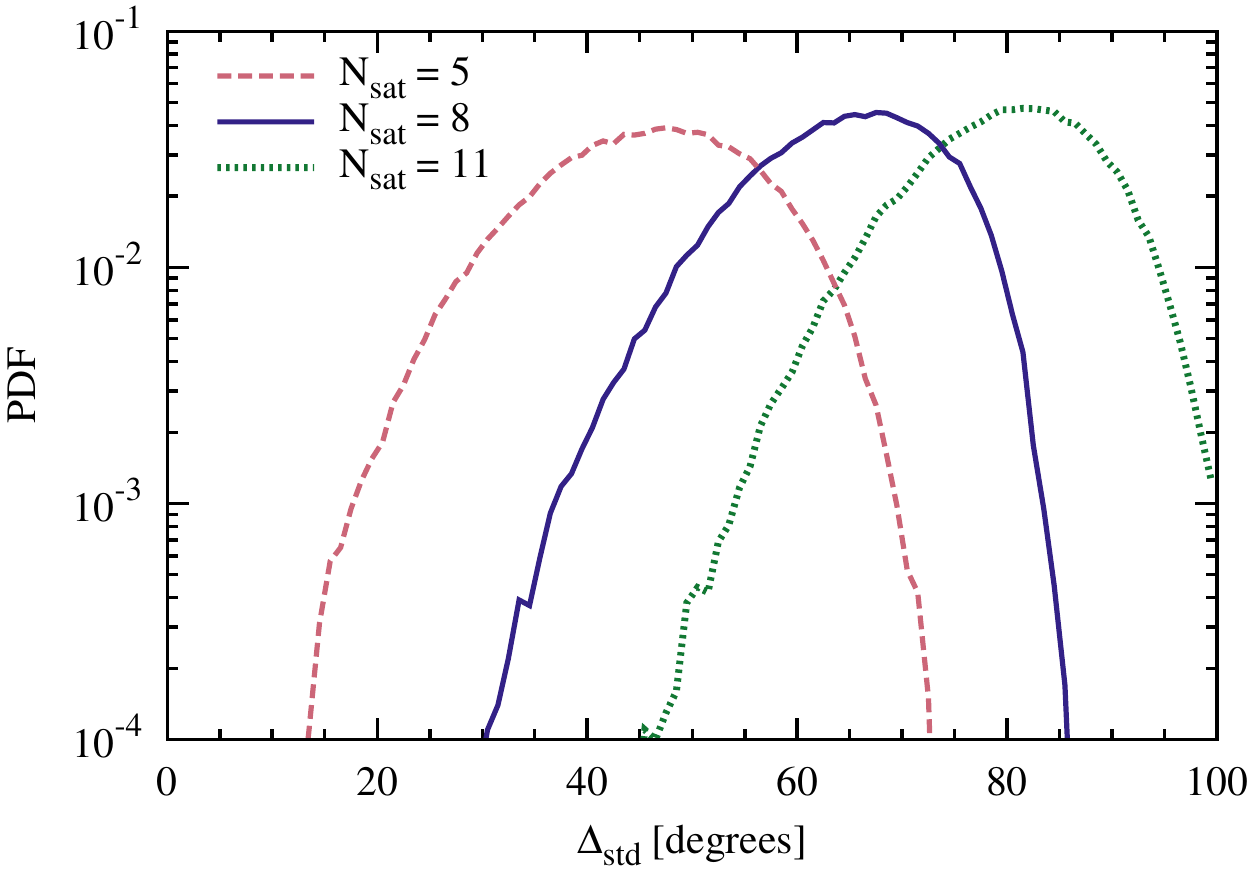}
     \caption{ The PDF of the angular dispersion of orbital poles, $\Dorbit$, for an isotropic satellite distribution. We show planar configurations that contain $\Nsat=5$, 8 and 11 satellites out of a maximum of $\Nmax=11$ satellites. To obtain these results, we used the same survey footprint and number of satellites as employed for the analysis of the MW satellite system in \refsec{sec:MW_planes}.}
     \label{fig:definition_orbital_width}
\end{figure}

The probability of obtaining by chance a plane with $\Nsat$ members that has an orbital pole dispersion less than $\Dorbit{}$ is calculated as
\begin{equation}
     p\left( \le \rPerp \;| \; \Nsat \right) = \int_{0}^{\Dorbit} PDF^{\rm isotropic}_{\rm 3D-kin;\; \Nsat}(\Delta'_{\rm std}) \;\mathrm{d} \Delta'_{\rm std}
    \label{eq:p_orbit_one_plane_2} \;.
\end{equation}
The integrand gives the PDF of the orbital pole dispersion, $\Delta'_{\rm std}$, for an isotropic distribution, which is estimated using $10^5$ random realizations, as for \eq{eq:p_spatial_one_plane_2}. For each such isotropic realization, we find the subsample of $\Nsat$ satellites that has the lowest orbital pole dispersion. The histogram of the lowest $\Delta'_{\rm std}$ values over all realizations gives the PDF used in \eq{eq:p_orbit_one_plane_2}. In \reffig{fig:definition_orbital_width} we illustrate the outcome of such a calculation for the case of $\Nsat=5$, 8 and 11 out of a maximum satellite count, $\Nmax=11$.

\end{document}